\documentclass{aa}
 \usepackage{graphicx,natbib}
 \bibpunct{(}{)}{;}{a}{}{,}
 \usepackage[figuresright]{rotating}
 
\newcommand{\EmEff}[1]{\mbox{$E_{#1}$}}
\newcommand{\ExtEmEff}[1]{\mbox{$\tilde{E}_{#1}$}}
\newcommand{\Signal}[1]{\mbox{$S_{#1}$}}
\newcommand{\Galaxy}[1]{\mbox{$G_{#1}$}}
\newcommand{\Zodiacal}[1]{\mbox{$Z_{#1}$}}
\newcommand{\Noise}[1]{\mbox{$N_{#1}$}}

\newcommand{\DeltaS}[1]{\mbox{$\Delta S_{#1}$}}
\newcommand{\DeltaG}[1]{\mbox{$\Delta G_{#1}$}}
\newcommand{\DeltaZ}[1]{\mbox{$\Delta Z_{#1}$}}
\newcommand{\DeltaN}[1]{\mbox{$\Delta N_{#1}$}}
\newcommand{\DeltaP}[1]{\mbox{$\Delta\mathbf{P}_{#1}$}}

\def\map{``map''}

\def\lsim{\,\lower2truept\hbox{${<\atop\hbox{\raise4truept\hbox{$\sim$}}}$}\,}
\def\gsim{\,\lower2truept\hbox{${>\atop\hbox{\raise4truept\hbox{$\sim$}}}$}\,}

\def\Planck{{\sc Planck}}

\def\Grad{\mbox{$\mathbf{\bar{\nabla}}$}}

\def\CalibRErr{\mbox{$\delta g$}}

\def\PErrSgm{\mbox{$\sigma_{\mathrm{p}}$}}

\def\Qabs{\mbox{$Q_{\mathrm{abs}}$}}

\def\Point{\mbox{$\mathbf{P}$}}

\def\SpaceCraft{\mbox{$\mathbf{R}_{\mathrm{P}}$}}
\def\ASpaceCraft{\mbox{$\overline{\mathbf{R}}_{\mathrm{P}}$}}
\def\DSpaceCraft{\mbox{$\delta \mathbf{R}_{\mathrm{P}}$}}
\def\DSpaceCrafti{\mbox{$\delta R_{\mathrm{P},i}$}}
\def\DSpaceCraftj{\mbox{$\delta R_{\mathrm{P},j}$}}
\def\DSpaceCraftk{\mbox{$\delta R_{\mathrm{P},k}$}}

\def\SpaceCraftLT{\mbox{$\mathbf{R}_{\mathrm{L2}}$}}
\def\DSpaceCraftLJ{\mbox{$\delta\mathbf{R}_{\mathrm{LJ}}$}}

\def\Sun{\mbox{$\mathbf{R}_{\odot}$}}
\def\ASun{\mbox{$\overline{\mathbf{R}}_{\odot}$}}
\def\DSun{\mbox{$\delta \mathbf{R}_{\odot}$}}
\def\DSuni{\mbox{$\delta R_{\odot,i}$}}
\def\DSunj{\mbox{$\delta R_{\odot,j}$}}

\def\Ef{\mbox{$E_f$}}

\def\LT{L$_2$}

 \def\FSZOD{{\tt FS\_ZOD}}
 \def\OCTAVE{{\tt OCTAVE}}
 
 \def\HEALpix{{\tt HEALPix}}

\def\alphaf{\mbox{$\alpha_{f}$}}
\def\ExtEf{\mbox{$\tilde{E}_{f}$}}
\def\Extalphaf{\mbox{$\tilde{\alpha}_{f}$}}

\def\ZodMod{\mbox{$Z$}}
\def\GalPat{\mbox{$G$}}

\def\SumSZ{\mbox{$\Sigma_{SZ}$}}

\def\SumZZ{\mbox{$\Sigma_{ZZ}$}}

\def\cov{\mbox{$\mathrm{cov}$}}
\def\var{\mbox{$\mathrm{var}$}}
\def\expect{\mbox{$\mathrm{E}$}}

\def\SumDSDZ{\mbox{$\Sigma_{\Delta S,\Delta Z}$}}

\def\SumDZDZ{\mbox{$\Sigma_{\Delta Z,\Delta Z}$}}

\def\Nside{\mbox{$N_{\mathrm{side}}$}}

\def\sigmaDN{\mbox{$\sigma_{\Delta N}$}}

\def\Kf{\mbox{$K_{f}$}}
\def\Wf{\mbox{$W_{f}$}}
\def\WfZ{\mbox{$W_{0,f}$}}

\def\densFf{\mbox{$\mathcal{F}_{f}$}}

\def\SSBar{\mbox{$B_{\mathrm{SS}}$}}
\def\ComponentCenter{\mbox{$C_{c}$}}

\def\DiscRadius{\mbox{$r_{\mathrm{dsc}}$}}
\def\CircleBeta{\mbox{$\beta$}}
\def\StepCircle{\mbox{$\delta\ell$}}

\def\tFirst{\mbox{$t_{\mathrm{I}}$}}
\def\tSecond{\mbox{$t_{\mathrm{II}}$}}

\def\Ncpl{\mbox{$N_{\mathrm{cpl}}$}}
\def\Nsmp{\mbox{$N_{\mathrm{smp}}$}}

\def\BeamSize{\mbox{$b$}}

\def\Gcut{\mbox{$G_{\mathrm{cut}}$}}
\def\sigmaEf{\mbox{$\sigma_{E_f}$}}
\def\ErrPatch{\mbox{$E_{\mathrm{r},2^\circ}$}}

\def\first{i)}
\def\second{ii)}
\def\third{iii)}
\def\fourth{iv)}
\def\fifth{v)}

 \def\TABEFRANDOMPOINTING{
 \begin{table}[t]
 \centering
 \begin{tabular}{ccc}
 \hline
 \hline
  $\sigma_{\mathrm{p}}$ & 
       \multicolumn{2}{c}{$\Delta E_{\mathrm{f}}$}  \\
  arcmin & $1^\circ$ Patches &   $2^\circ$ Patches \\
 \hline
 0.5 &  $(6.3 \pm 18.4)\times10^{-5}$ & $(1.0 \pm 2.09)\times10^{-4}$ \\
 1.0 &  $(8.7 \pm 27.9)\times10^{-5}$ & $(0.9 \pm 3.13)\times10^{-4}$ \\
 1.5 &  $(5.2 \pm 4.10)\times10^{-5}$ & $(1.7 \pm 4.50)\times10^{-4}$ \\
 \hline
 \hline
 \end{tabular}
 \caption{
 Expected effect at 857~GHz of an isotropic, random pointing error on 
 the \Ef\ sensitivity as a function of $\PErrSgm$. Calculations are for 
 a cut of 1~MJy/sr on the Galaxy and $1^\circ$ and $2^\circ$ patches. 
 }\label{Tab:Ef:Random:Pointing} 
 \end{table}
}

\def\TABLEEFFD{
 \begin{table}
 \begin{center}
 \begin{tabular}{rccc}
 \hline
 \multicolumn{1}{c}{f} 
 &
 \multicolumn{1}{c}{$\langle I_{\nu=f}\rangle_{\mathrm{year}}^{\mathrm{FD}}$}
 &
 \multicolumn{1}{c}{$\langle Z_{f} \rangle_{\mathrm{year}}$}
 &
 \multicolumn{1}{c}{$E_f^{\mathrm{FD}}$}
 \\
 \multicolumn{1}{c}{[GHz]} 
 &
 \multicolumn{1}{c}{[MJy/sr]} 
 &
 \multicolumn{1}{c}{[MJy/sr]} 
 &
 \\
 \hline
 30 & 0.0000 & 0.0006 & 0.001 \\
 44 & 0.0000 & 0.0013 & 0.002 \\
 70 & 0.0000 & 0.0032 & 0.004 \\
100 & 0.0001 & 0.0064 & 0.009 \\
143 & 0.0002 & 0.0129 & 0.018 \\
217 & 0.0012 & 0.0291 & 0.041 \\
354 & 0.0083 & 0.0755 & 0.110 \\
545 & 0.0458 & 0.1751 & 0.262 \\
857 & 0.2742 & 0.4229 & 0.648 \\
 \hline
 \end{tabular}
 \end{center}
 \caption{
$E_{f, \mathrm{Smooth}}^{\mathrm{FD}}$ estimated
 according to Eq.~(\ref{eq:Ef:FD}).
 }\label{tab:Ef:FD}
 \end{table}
 } 

\def\TABEFUNCONEDEG{
 \begin{table*}
 \centering
 \begin{tabular}{ccccccccc}
 \multicolumn{1}{c}{$G\le G_{\mathrm{cut}}$} &
 \multicolumn{1}{c}{$N_{\mathrm{dsc}}$} &
 \multicolumn{1}{c}{$\cov(Z,G)$} &
 \multicolumn{1}{c}{$\var(Z)$} &
 \multicolumn{1}{c}{$\var(G)$} &
 \multicolumn{1}{c}{$\frac{\cov(Z,G)}{\var(Z)}$} &
 \multicolumn{1}{c}{$\cov(\Delta Z,G)$} &
 \multicolumn{1}{c}{$\var(\Delta Z)$} &
 \multicolumn{1}{c}{$\sigma_{E_f}$} 
 \\
 \multicolumn{1}{c}{[MJy/sr]} &
   &
 \multicolumn{1}{c}{[MJy$^2$/sr$^2$]} &
 \multicolumn{1}{c}{[MJy$^2$/sr$^2$]} &
 \multicolumn{1}{c}{[MJy$^2$/sr$^2$]} &
   &
 \multicolumn{1}{c}{[MJy$^2$/sr$^2$]} &
 \multicolumn{1}{c}{[MJy$^2$/sr$^2$]} &
 \\
 \hline
 $\infty$&    8010&    1.60E-1&    4.72E-2&    1.12E+3&       3.4&     6.25E-2&     1.87E-3&       6.9E-4\\ 
      4.0&    5141&    2.01E-2&    4.62E-2&    9.76E-1&      0.43&     7.57E-3&     1.49E-3&       9.8E-4\\ 
      3.0&    4524&    2.38E-2&    4.55E-2&    5.29E-1&      0.52&     4.75E-3&     1.40E-3&       1.1E-3\\ 
      2.0&    3591&    1.52E-2&    4.38E-2&    2.12E-1&      0.35&     8.10E-4&     1.18E-3&       1.3E-3\\ 
      1.0&    2007&    7.58E-3&    3.27E-2&    3.49E-2&      0.23&    -3.94E-4&     9.09E-4&       2.0E-3\\ 
      0.9&    1772&    7.20E-3&    2.95E-2&    2.56E-2&      0.24&    -4.56E-4&     8.67E-4&       2.1E-3\\ 
      0.8&    1495&    5.42E-3&    2.48E-2&    1.76E-2&      0.22&    -3.27E-4&     8.09E-4&       2.4E-3\\ 
      0.7&    1220&    3.98E-3&    1.95E-2&    1.17E-2&      0.20&    -1.91E-4&     7.68E-4&       2.8E-3\\ 
      0.6&     881&    2.32E-3&    1.44E-2&    7.25E-3&      0.16&    -8.37E-5&     6.96E-4&       3.4E-3\\ 
      0.5&     505&    1.54E-3&    1.06E-2&    4.17E-3&      0.14&     8.33E-5&     6.67E-4&       4.7E-3\\ 
      0.4&     177&    7.58E-4&    5.51E-3&    2.39E-3&      0.14&     9.41E-5&     5.72E-4&       8.5E-3\\ 
 \hline
 \end{tabular}
 \caption{
 Correlation between Galaxy, ZLE and Differential ZLE 
 at 857~GHz,
 for 
   $\DiscRadius =     1^\circ$,
   $\CircleBeta =      85^\circ$,
   $\StepCircle =      2^\circ$.
 }\label{tab:gal:zle:corr}
 \end{table*}
} 

\def\TABEFCALIBERR{
 \begin{table*}[t]
 \centering
 \begin{tabular}{ccccc}
 \hline
 \hline
 \multicolumn{1}{c}{Freq} &
 \multicolumn{1}{c}{Minimum $\Delta \Ef$} &
 \multicolumn{1}{c}{Optimal \Gcut} &
 \multicolumn{1}{c}{Min. \Gcut} &
 \multicolumn{1}{c}{Max. \Gcut} 
 \\
 \multicolumn{1}{c}{[GHz]} &
 \multicolumn{1}{c}{} &
 \multicolumn{1}{c}{[MJy/sr]} &
 \multicolumn{1}{c}{[MJy/sr]} &
 \multicolumn{1}{c}{[MJy/sr]} 
 \\
 \hline
 857 & 0.022 & 6.5 & 2.3 & 26 \\
 545 & 0.017 & 2.0 & 1.0 & 8.3 \\
 353 & 0.011 & 0.50 & 0.20 & 2.0 \\
 \hline
 \hline
 \end{tabular}
 \caption{
 Effect of a random relative calibration error 
 $\ErrPatch = 1\%$ 
 on the \Ef\ determination at different \Planck\ frequencies.  
 Column 2 is the minimum RMS error for the optimal surface brightness cut 
 reported in Column 3. 
 Columns 4 and 5 give the range for which the RMS error is 
 less than twice the value in Column 2.
 }\label{Tab:Ef:Calibration:Error} 
 \end{table*}
} 

\def\ColorFigureStat{(Colour figure on the electronic version.)}

\def\FIGSCANGEOM{
 \begin{figure}
 \centering
\resizebox{\hsize}{!}{\includegraphics{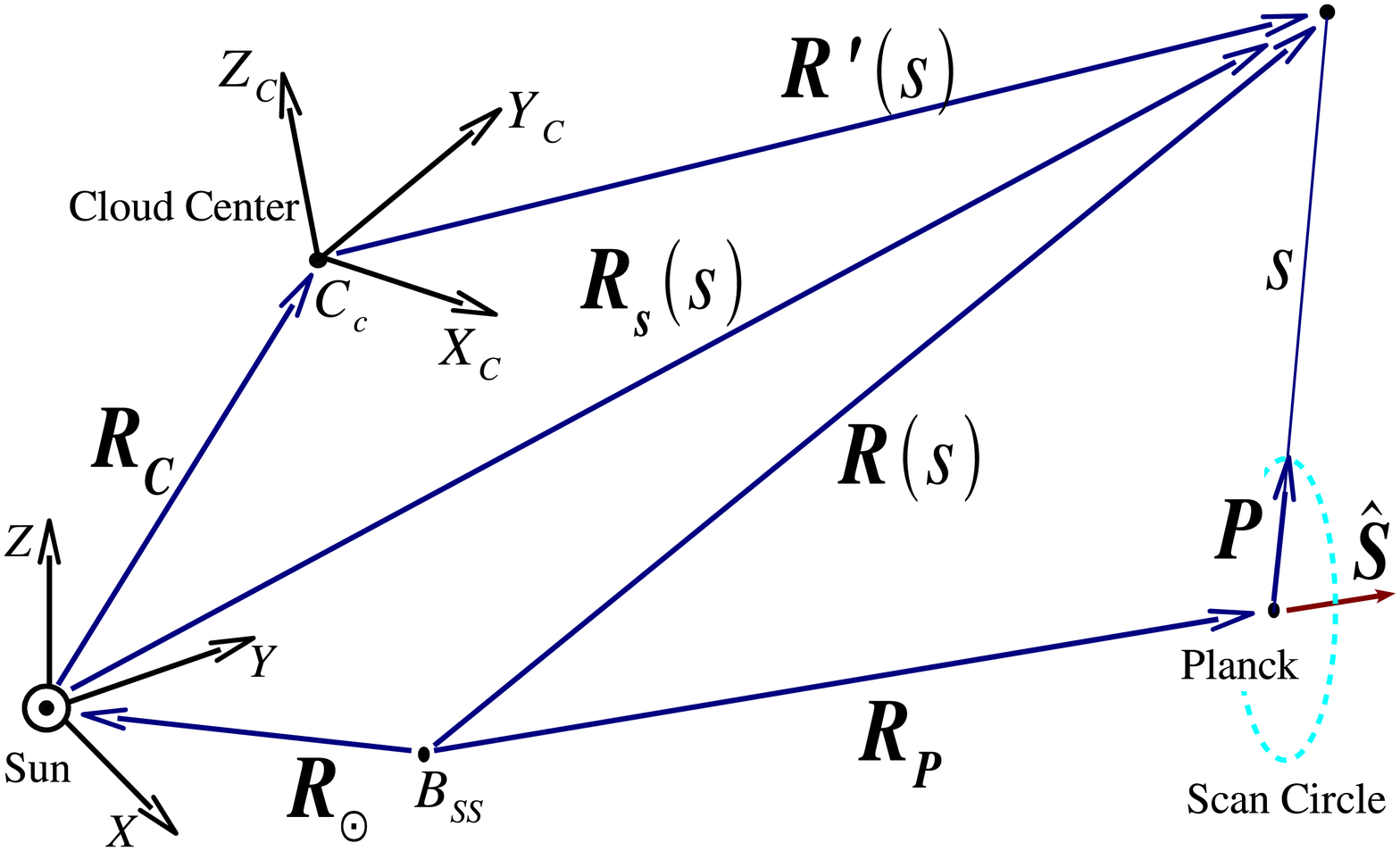}}
 \caption{
Relations between the heliocentric and cloud-centred frames used to 
describe the Solar System scanning geometry for a mission like 
{\sc Planck}. 
The relations between Planck, Sun, Cloud Center and 
the observed portion of cloud are drawn, 
the \LT\ point is not drawn to simplify the graph.
The connection between the other two reference frames is shown in
Fig.~\ref{fig:scanning:geometry:b}.
An example of scan circle and the related spin axis, $\hat{\mathbf{S}}$,
are also drawn.
The graph is not in scale with real distances.
\ColorFigureStat
  }\label{fig:scanning:geometry}
 \end{figure}
} 

\def\FIGSCANGEOMB{
 \begin{figure}
 \centering
\includegraphics[width=\hsize]{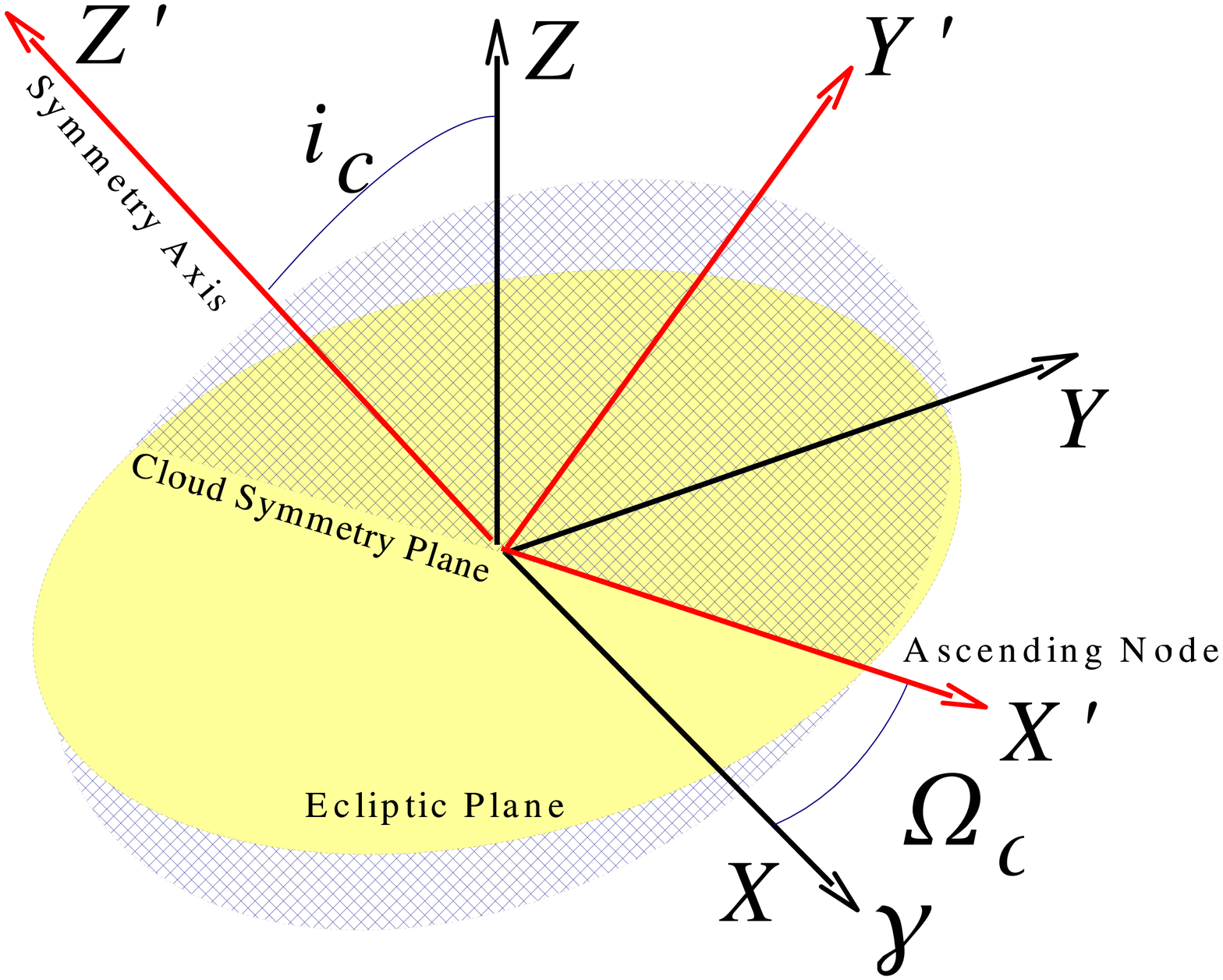}
\resizebox{8cm}{!}{\includegraphics{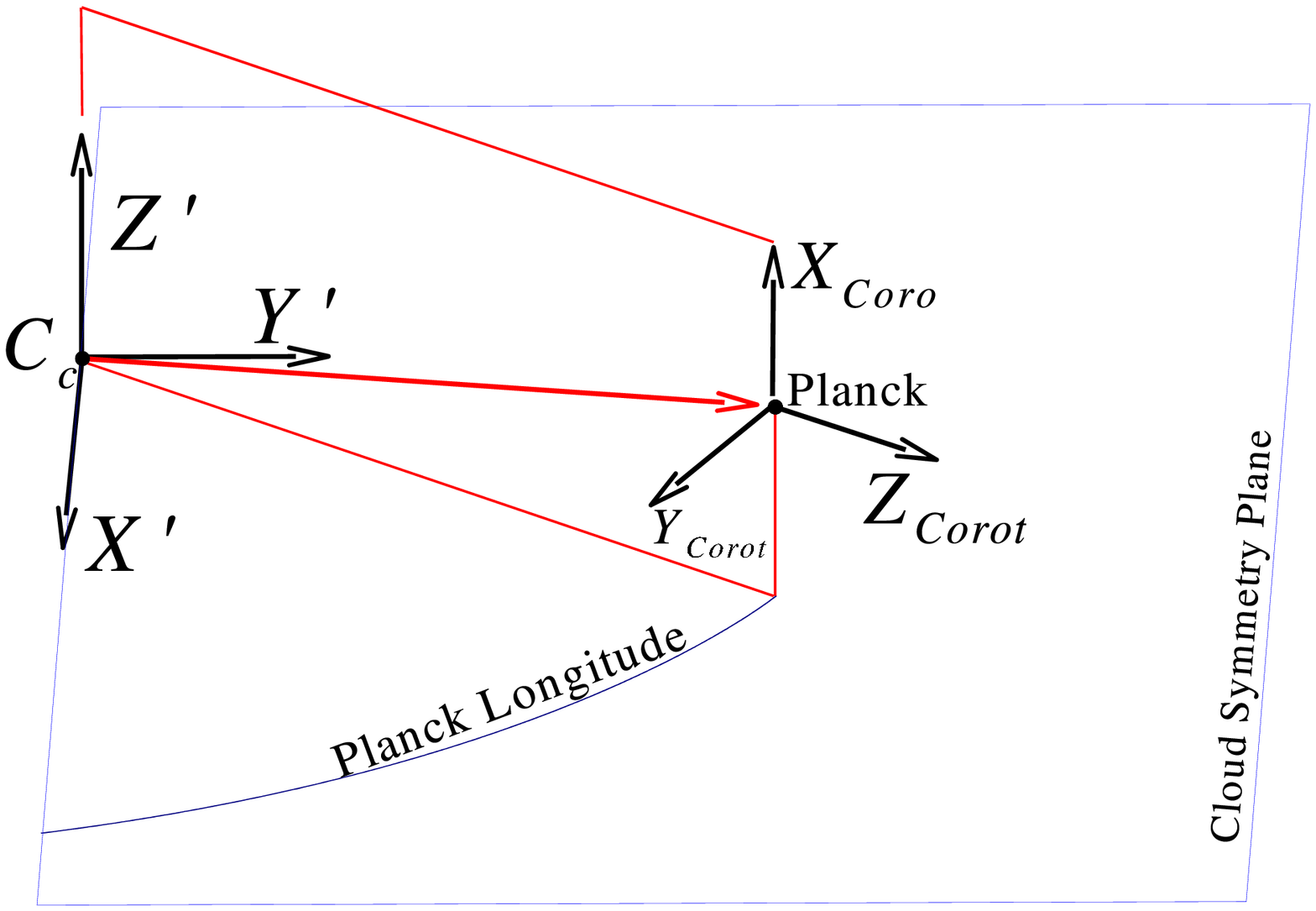}}
\vskip -3cm
 \caption{
Relations between the cloud-centred ecliptic reference frame,
the cloud symmetry reference frame and the spacecraft 
centred corotating reference frame. 
Top panel: angles between the cloud-centred Ecliptical reference 
frame (full yellow disc (light-gray in the bw version) - black arrows), 
and the cloud cylindrical symmetry 
reference frame (dashed blue disc - red arrows).
Bottom panel: angles between the cloud cylindrical symmetry reference 
frame, the cloud-centred corotating cylindrical reference frame,
the spacecraft - centred corotating reference frame.
The displacement of \Planck\
(in this case above the symmetry plane), 
\DSpaceCraft,
respect to the constant reference position, 
 \ASpaceCraft, assumed to be in the symmetry plane
of the corotating reference frame is also displayed.
The graph is not in scale with real distances.
\ColorFigureStat
  }\label{fig:scanning:geometry:b}
 \end{figure}
} 

 \def\FIGSMOOTHCNT{
  \begin{figure}
  \centering
  \resizebox{\hsize}{!}{\rotatebox{+90}{\includegraphics{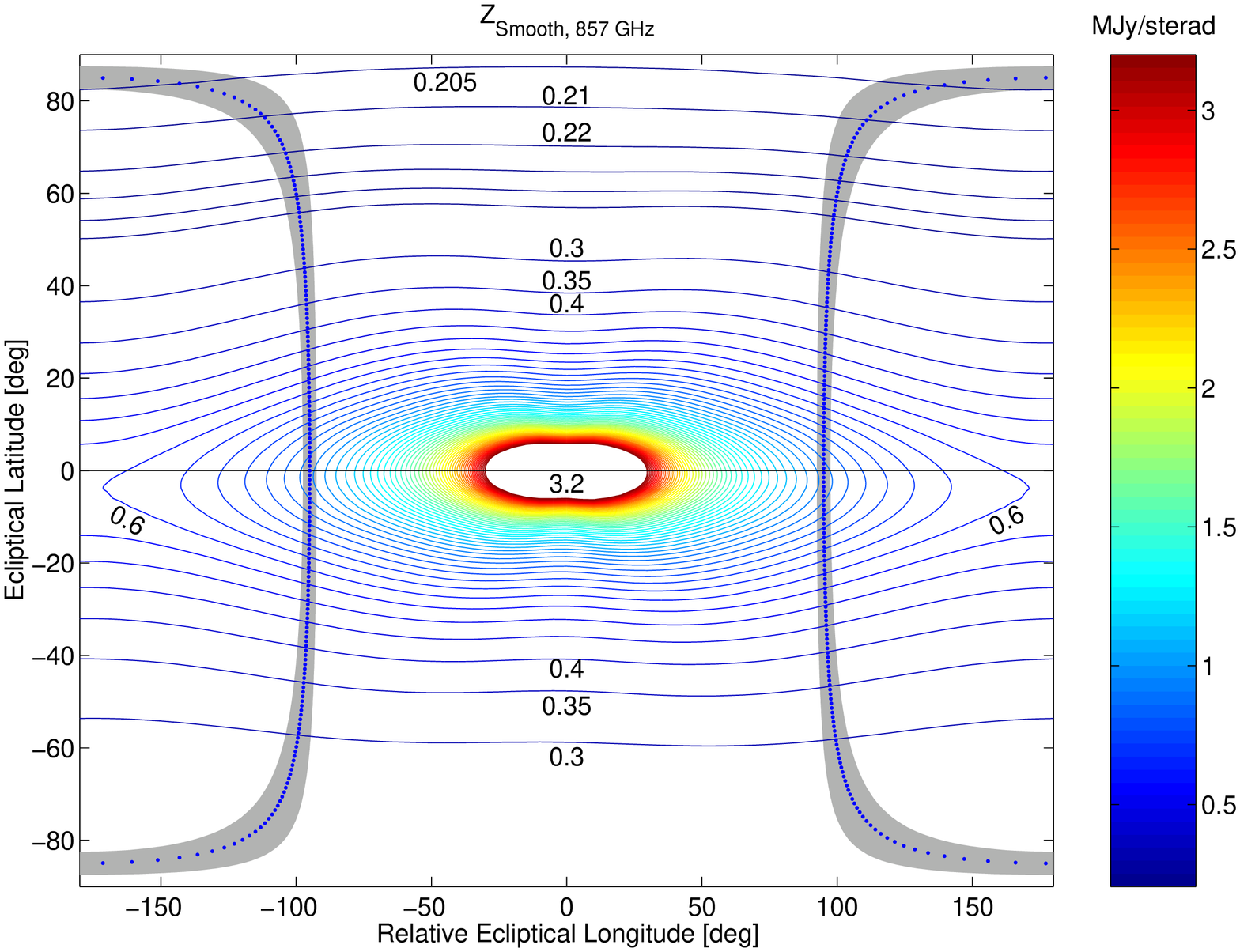}}}
 \caption{
Simulated contour plot of $Z_f(\Point)$~[MJy/sr] for a fixed
location within the Solar System. Contours are calculated for $Z_f =
0.205$, 0.22, 0.23, 0.24~MJy/sr  and from 0.25~MJy/sr up
to 3.2~MJy/sr in steps of $0.05$~MJy/sr. 
Surface brightnesses exceeding 3.2~MJy/sr 
occurring in the central empty region have 
been neglected.
The blue-dotted
line represents the path of a scan circle of {\sc Planck} for an
horn located at the centre of the field of view while the gray band
represents the {\sc Planck} field of view. The vertical axis is
the latitude over the ecliptic. The horizontal axis the longitude
relative to the solar direction (opposite to orientation of 
the scan axis). 
Note that in this case the symmetry plane of the IDP cloud is below
the ecliptic plane.
\ColorFigureStat
  }\label{Fig:Contour:Smooth}
   \end{figure}
 } 

 \def\FIGONE{
   \begin{figure}
 \centering
 \resizebox{7cm}{!}{\rotatebox{+90}{\includegraphics{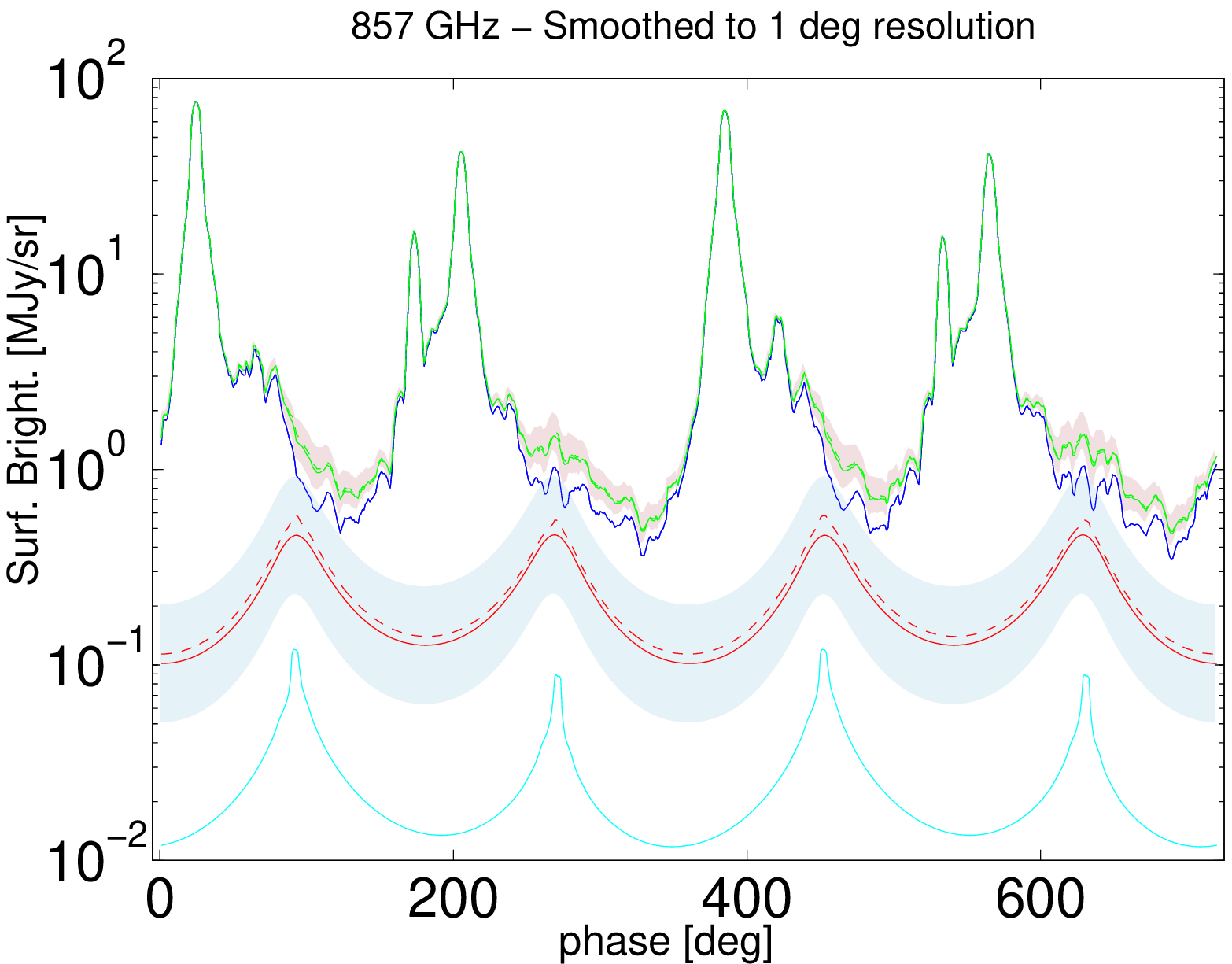}}}
 \caption{
Simulated data stream of surface brightnesses (MJy/sr) measured at
857~GHz for the ZLE - smooth component (red), the Galaxy (green)
and the sum of the two (blue).
The ordinate is the phase of the scan circle, assumed to be zero for the pointing direction
nearest to the North ecliptic Pole. Two subsequent scan circles are displayed, the phase
of the second being augmented of $360^\circ$.
The lowest, full, cyan line (peaks at 0.1 MJy/sr) represents the contribution
of secondary components of the ZLE. The
red-dashed line represents the full contribution of ZLE with both smooth
component and secondary components.
The cyan band represents the uncertainty in the prediction for the
Smooth component due to the uncertainty in the extrapolated \Ef\ value in the case
of $m_c \gsim 10^{-6}$~gr.
Both the dashed line and the cyan band are reported after adding the Galactic
contribution as a green dashed line and a grey band above the Galaxy.
\ColorFigureStat
 }\label{Fig1}
   \end{figure}
 } 

\def\FIGYEARLYAVER{
 \begin{figure}
 \centering
\resizebox{\hsize}{!}{\rotatebox{+90}{\includegraphics{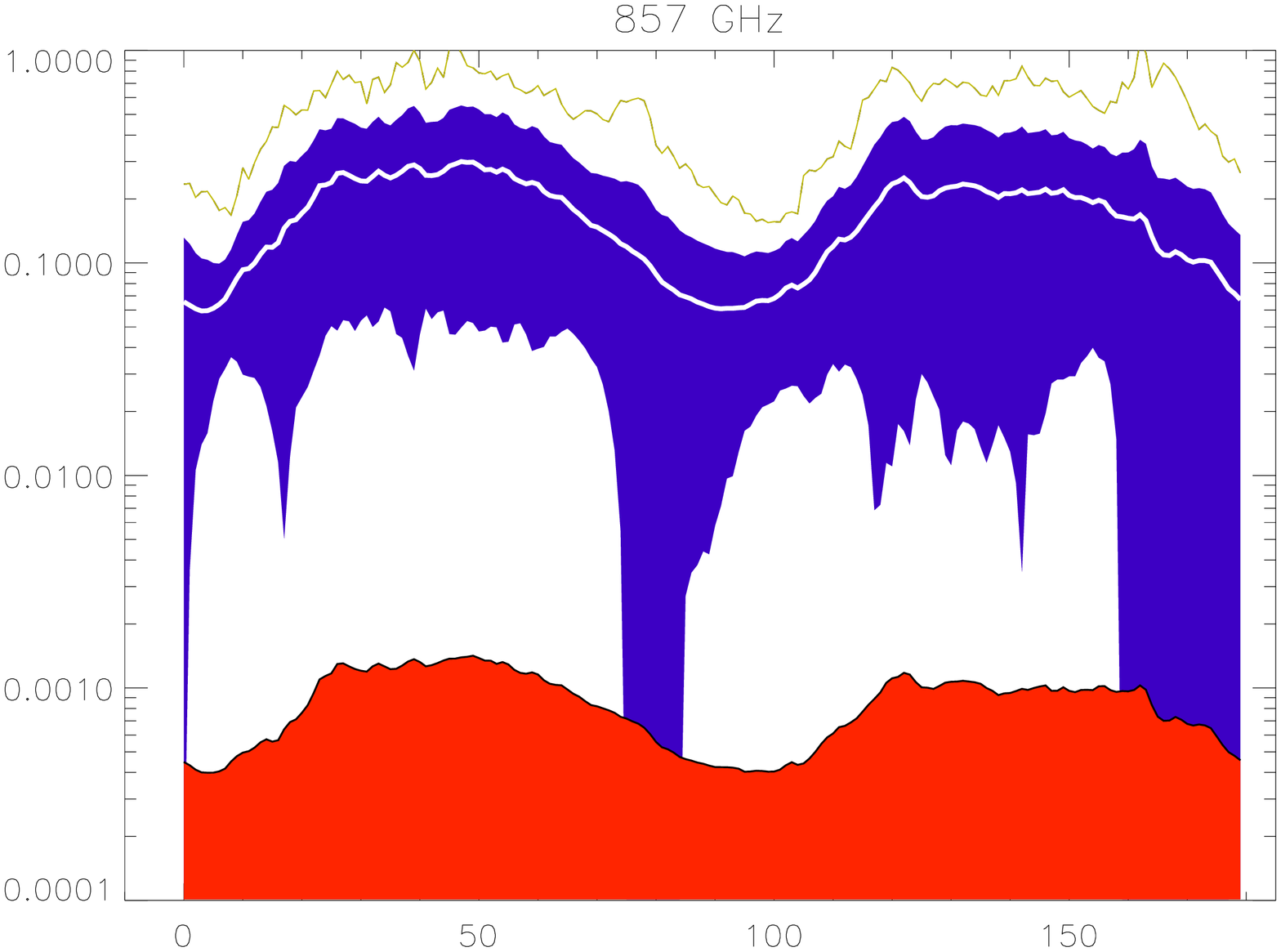}}}
\resizebox{\hsize}{!}{\rotatebox{+90}{\includegraphics{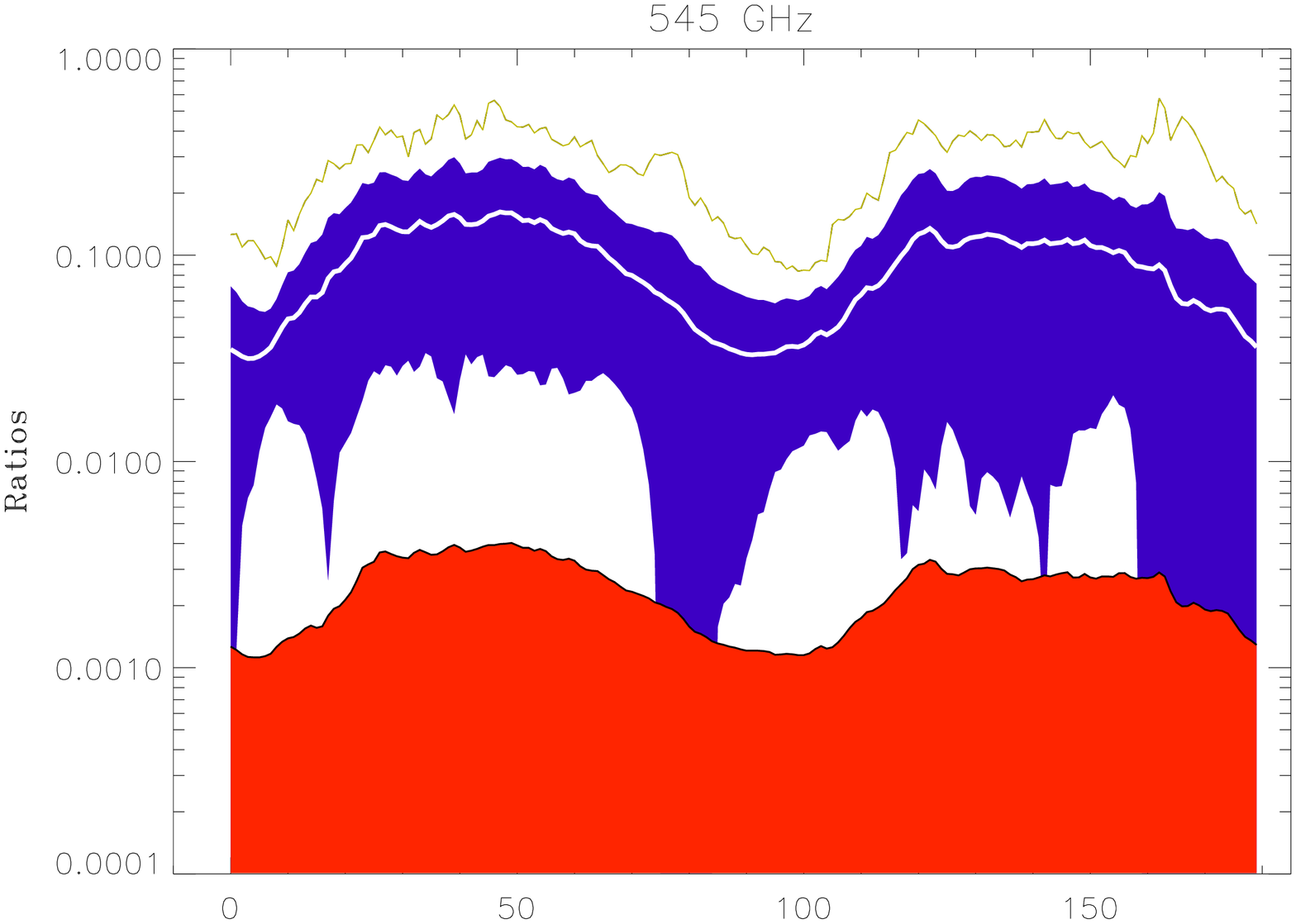}}}
\resizebox{\hsize}{!}{\rotatebox{+90}{\includegraphics{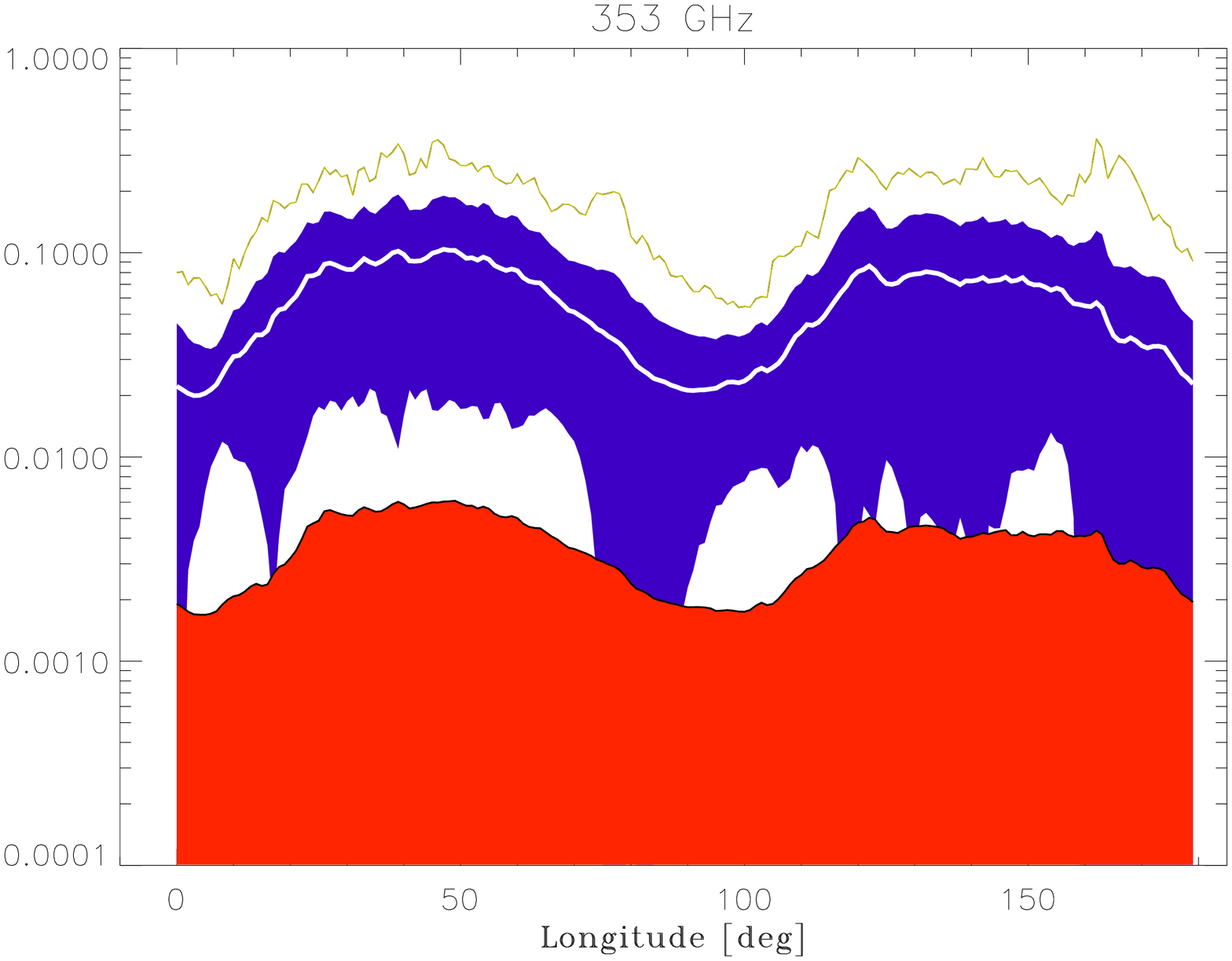}}}
 \caption{
Relative contribution of ZLE, Galactic (dust) emission and noise as a function
of the pointing ecliptical longitude at three \Planck\ frequency bands: 857~GHz (top), 
545~GHz (middle), 343~GHz (bottom).
For each pointing ecliptical longitude, the plots gaves the ratio of the
ZLE over the Galactic emission $\Ef Z_f/G_f$ averaged over the given scan circle 
(white - full line), the $\pm3\sigma$ range 
(blue (black in the bw version) band), the peak ratio (yellow (gray) - full line) and 
the averaged ratio between the instrumental noise and the Galactic emission. 
Data are calculated for patches of $1^\circ$ in radius, noise is for a 14 month mission 
(2 sky surveys).
\ColorFigureStat
  }\label{fig:yearly:aver}
 \end{figure}
  
} 

 \def\FIGALONGACROSS{
  \begin{figure}
 \centering
 \resizebox{\hsize}{!}{\rotatebox{+90}{\includegraphics{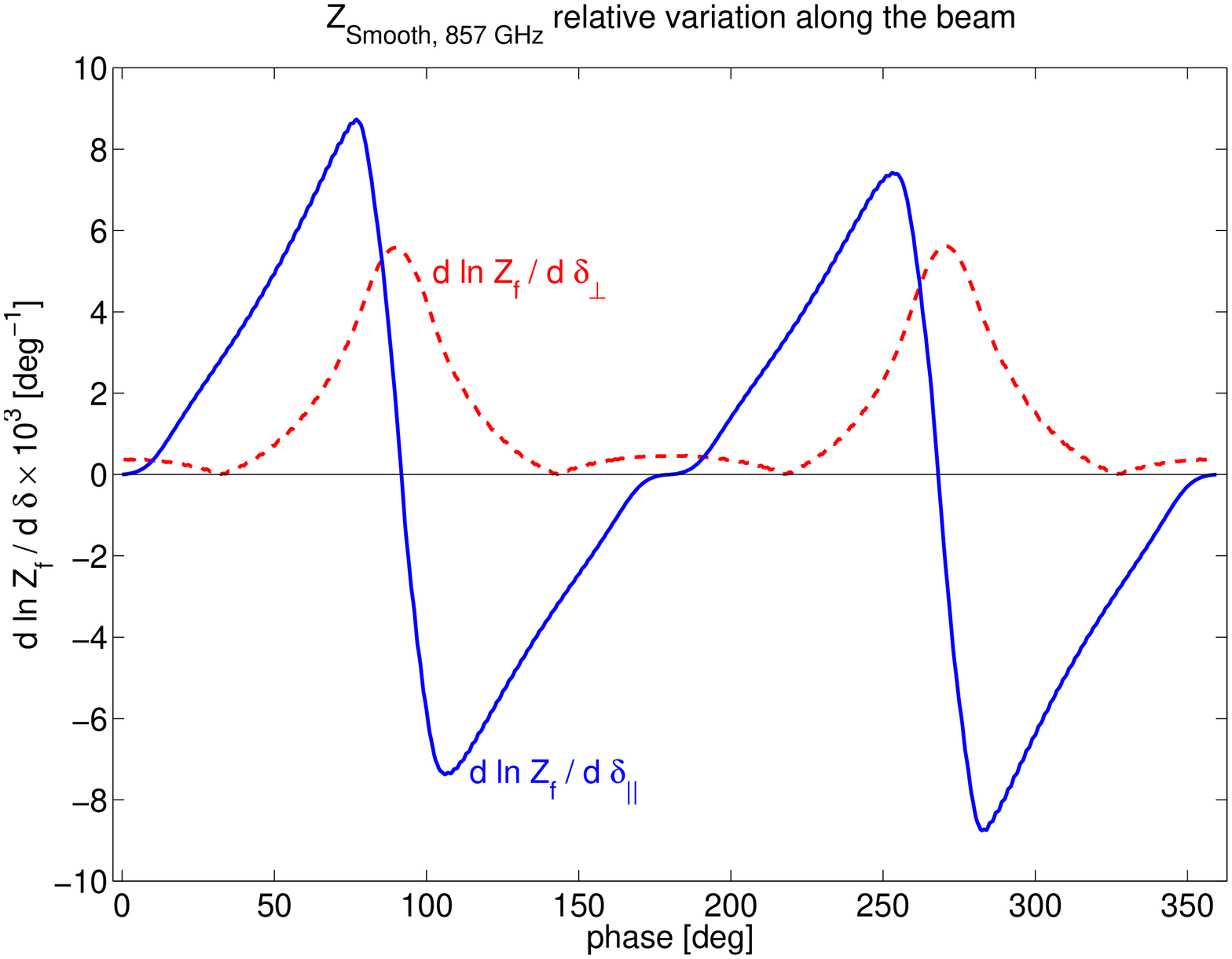}}}
 \caption{
Derivatives of $Z_{\mathrm{Smooth}, 857~\mathrm{GHz}}$ along (blue full line - $\delta_\parallel$) 
and across (red dashed line - $\delta_\perp$) the scan circle of Fig.~\ref{Fig:Contour:Smooth}.
For graphical reasons derivatives in the plot are scaled by a factor $10^3$.
According to the convention of Fig.~\ref{Fig1} the phase represents the pointing position along 
the scan circle, with 0 for the pointing direction nearest to the North ecliptical Pole.
Note that the $\delta_\perp$ direction is always normal to $\delta_\parallel$ and oriented
toward the local solar direction so that $d\ln Z_f/d \,\delta_\perp$ is always positive.
\ColorFigureStat
 }\label{fig:along:across}
   \end{figure}
 } 

\def\FIGYEARLYAVERCIRC{
 \begin{figure}
 \centering
\resizebox{\hsize}{!}{
 \rotatebox{+90}{
\includegraphics{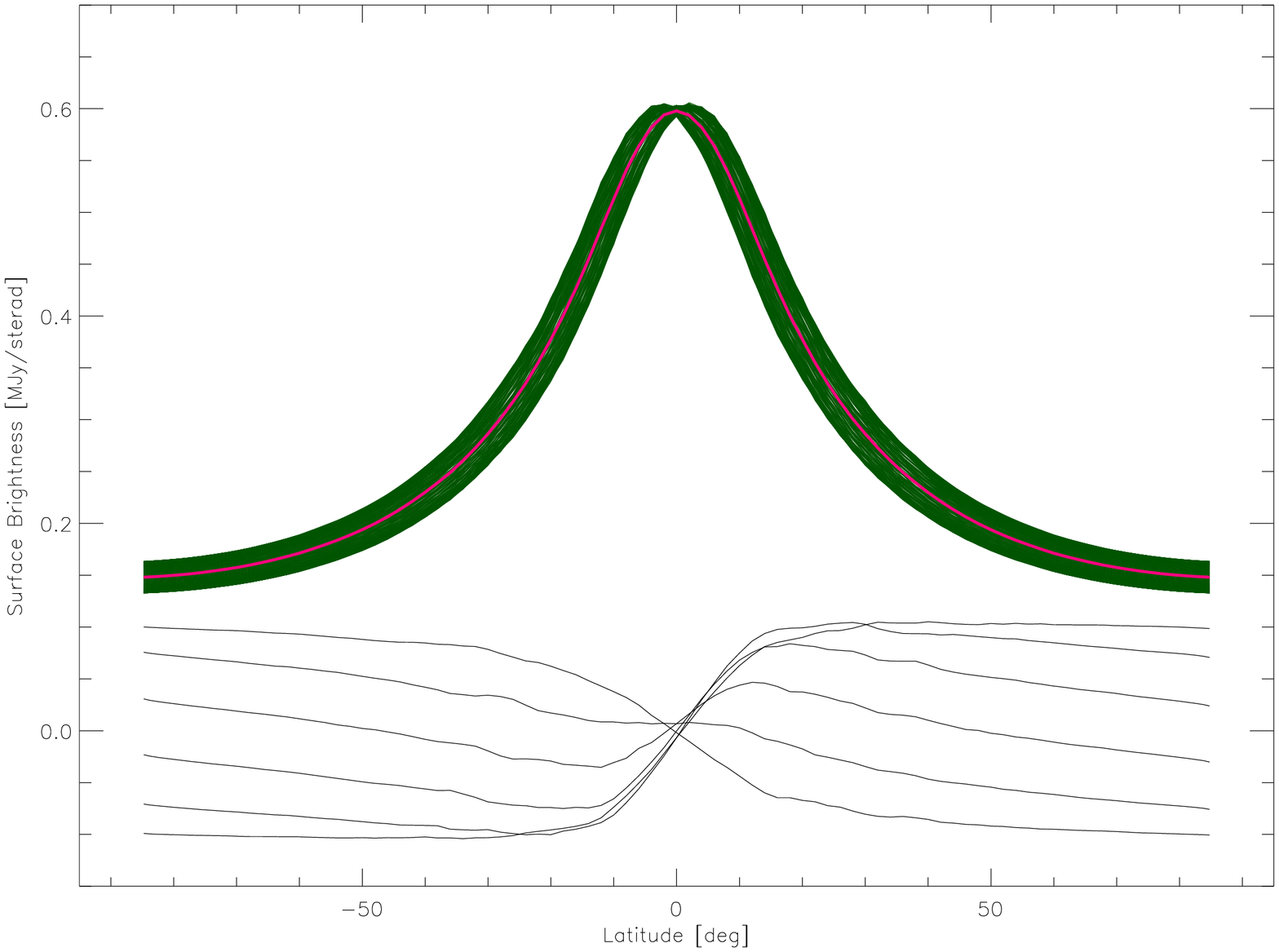}
 }}
 \caption{
Absolute and relative variation of the ZLE surface brightness during the 
year at 857~GHz. 
The full-red line (light-gray in bw) is the yearly averaged ZLE surface brightness 
[MJy/sr] for a given eclipitical latitude.
The surrounding green band (dark-gray in bw) is the variation of ZLE surface brightness 
during the year.
Below (full black lines) the relative variation (with respect to the average)
of ZLE for a set of representative longitudes (from the left of the 
black curves, from $0^\circ$ to $75^\circ$ in steps of $15^\circ$, respectively
from top to bottom). Note that these relative variations reach peaks 
of $\simeq10\%$.
Data are calculated for patches of $1^\circ$ in radius, $E_{857\;\mathrm{GHz}} = 0.65$.
  }\label{fig:yearly:aver:circ}
 \end{figure}
} 

\def\FIGODDEVENASIM{
 \begin{figure}
 \centering
 \resizebox{\hsize}{!}{\rotatebox{+90}{\includegraphics{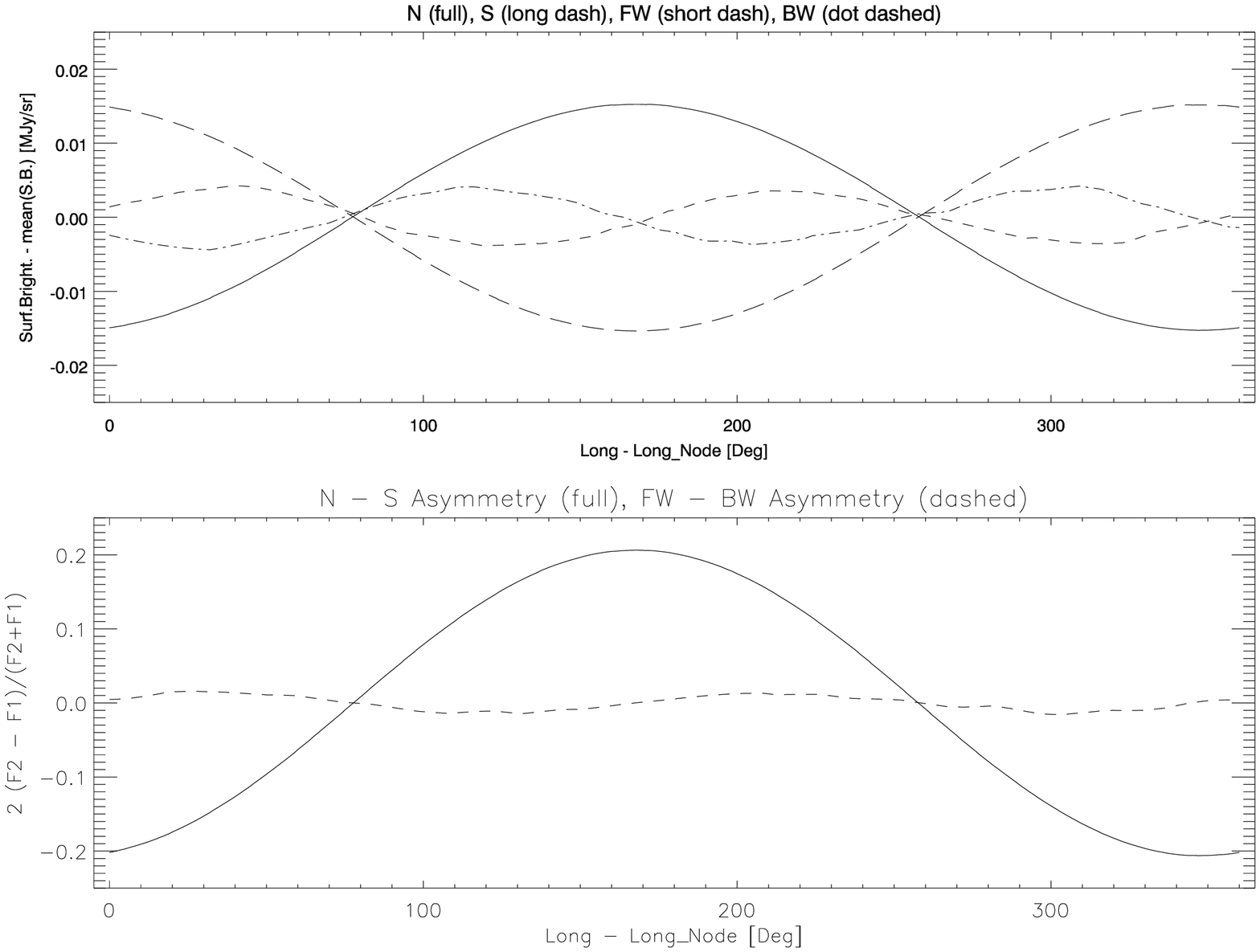}}}
 \caption{
 Seasonal modulation of the Smooth component of the ZLE at 857~GHz
 for four selected pointing directions: 
 the North ecliptic Pole (N), 
 the South eclitptic Pole (S), 
 the Forward direction with respect to the \Planck\ motion (FW) and 
 the Backward direction (BW).
 Longitudes are relative to the longitude of the ascending node.
 Upper frame: the surface brightness variation looking to N (full line), S 
(long dashed line), 
 FW (short dashed line),  BW (dot-dashed line). The variation is the difference 
between the surface brightness 
 along the
 direction of choice and its yearly average. Here $E_{\mathrm{Smooth}, \,857\,\mathrm{GHz}} = 0.65$.
 Lower frame: North - South (full line) and Forward - Backward (dashed line) asymmetries. 
 Asymmetries are defined as $A = 2(F_2 - F_1)/(F_2 + F_1)$ with $F_2 = F_{\mathrm{N}}$,
 $F_1 = F_{\mathrm{S}}$ for the North - South asymmetry and 
 $F_2 = F_{\mathrm{FW}}$,
 $F_1 = F_{\mathrm{BW}}$ in the other case.
  }\label{fig:odd:even:asim}
 \end{figure}
} 

\def\FIGBETAPLOT{
  \begin{figure}
 \centering
 \resizebox{\hsize}{!}{\rotatebox{+90}{\includegraphics{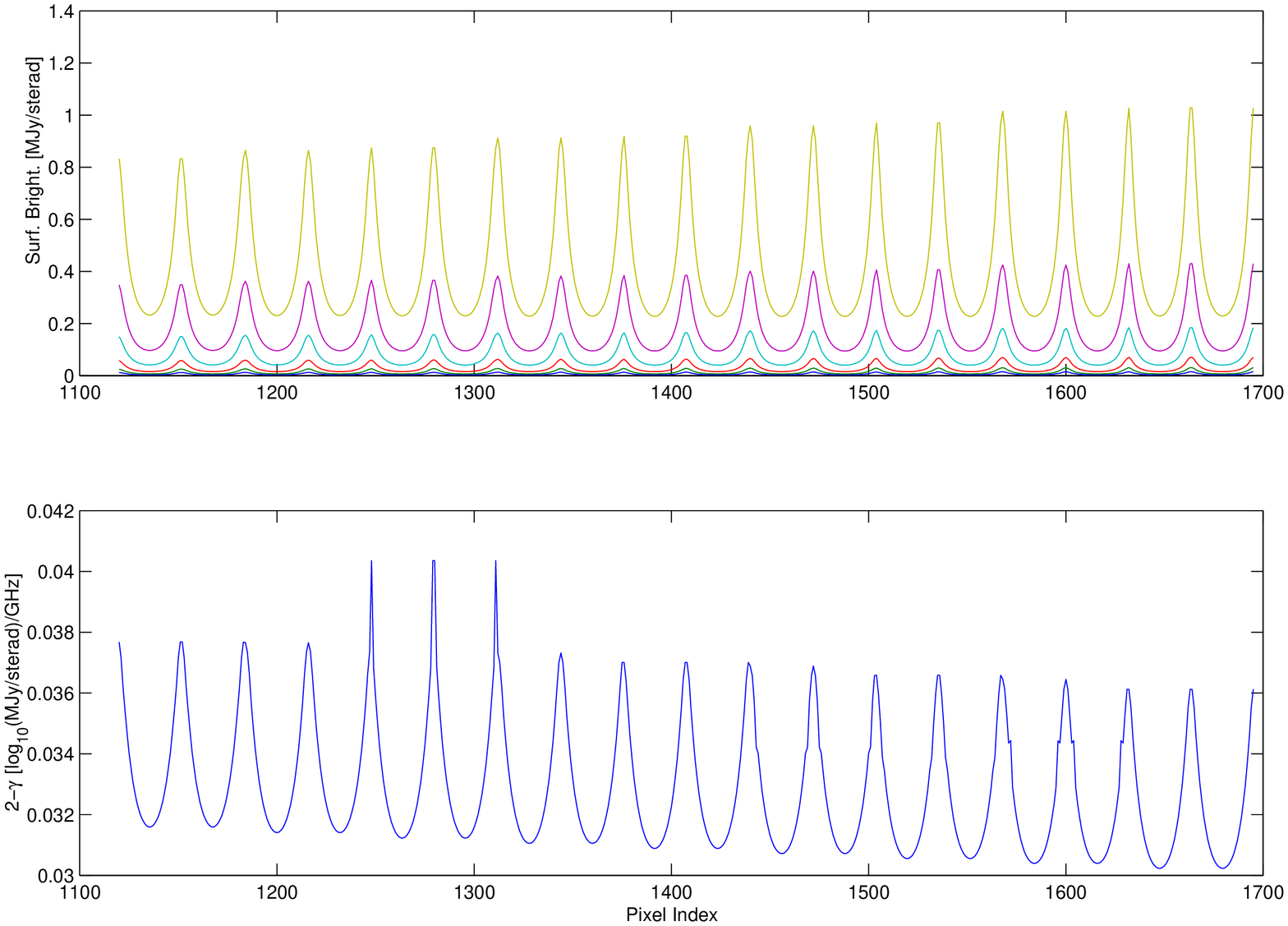}}}
 \caption{
Deviation of the ZLE from an $f^2$ scaling as a function of
the direction in the sky.
Upper frame: the original data for $f = 100$, 143, 217,
353, 545, 857~GHz (from lower to upper surface brightness).
Lower frame: deviation from the $f^2$ scaling in terms
of $2-\gamma_{f,z}$, where $\gamma_{f,z}$ is obtained by fitting a power law scaling on
the data in the upper frame.
In the horizontal axis we report the pixel indices for an {\tt HEALPix} map
with $\Nside = 16$
ordered according to the {\em ring} scheme. The colatitude in the
map decreases from left to right.
\ColorFigureStat
 }\label{Fig:Beta:Plot}
   \end{figure} 
} 

\def\FIGZZEROCOEF{
 \begin{figure}
 \centering
 \resizebox{\hsize}{!}{\rotatebox{-90}{\includegraphics{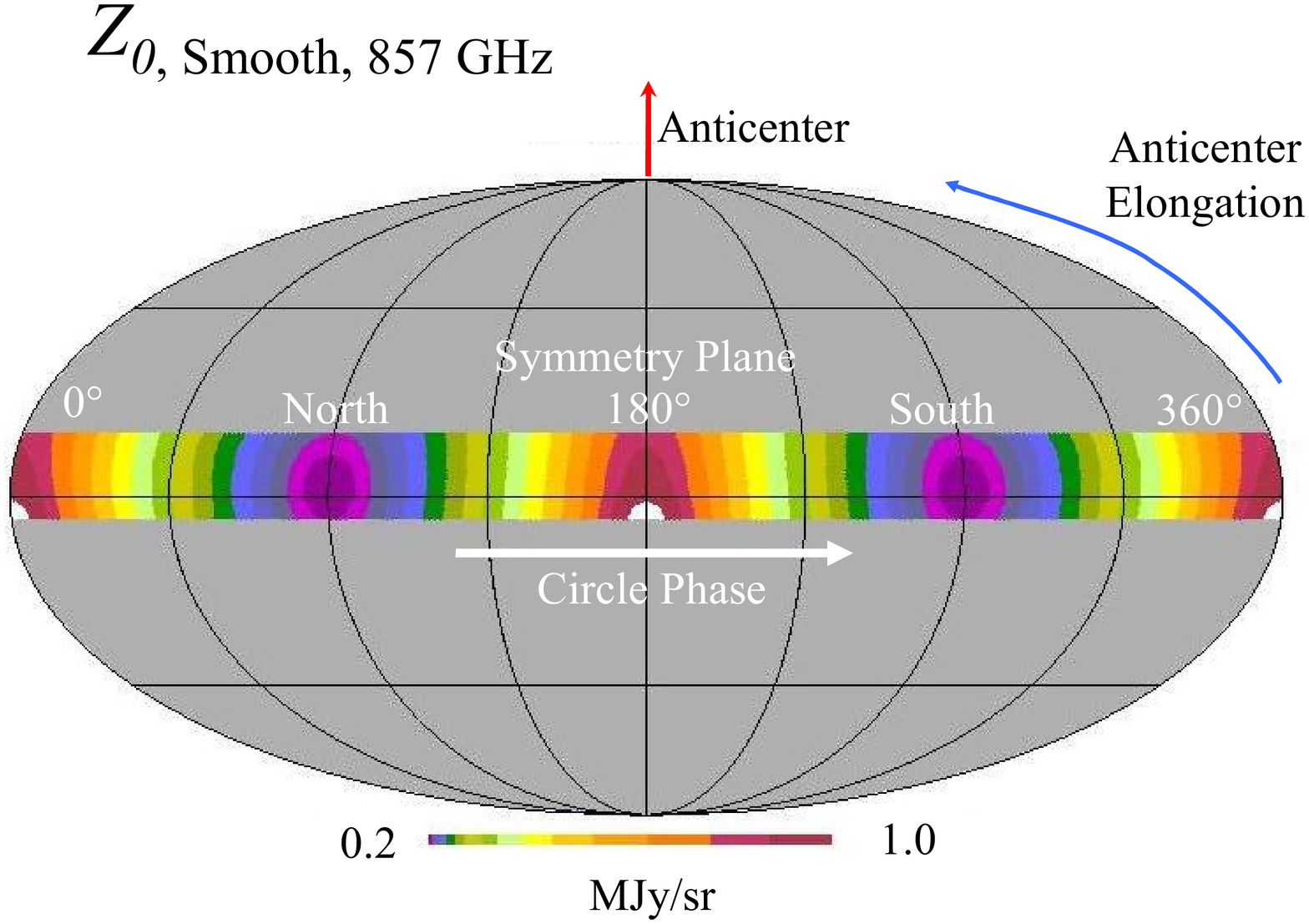}}}
 \caption{
Map of $\bar{Z}_{\mathrm{Smooth}, \; 857\;\mathrm{GHz}}$ with 
$\Nside = 128$. The top of the map corresponds to the direction 
of the anticenter of the cloud. The sides and the middle of the map to 
poiting directions in the cloud symmetry plane. 
Compared to
Fig.~\ref{Fig:Contour:Smooth}, the circles phase is 
 $0^\circ$ when the pointing direction lies on the equator near 
 $X_{\mathrm{corot}}$ and
 $90^\circ$ on the equator near $Y_{\mathrm{corot}}$
 (the North pole of the cloud). 
\ColorFigureStat
 }\label{fig:zzero:coef}
 \end{figure}
} 

\def\FIGZLEDIFF{
 \begin{figure}
 \centering
 \resizebox{\hsize}{!}{\rotatebox{+90}{\includegraphics{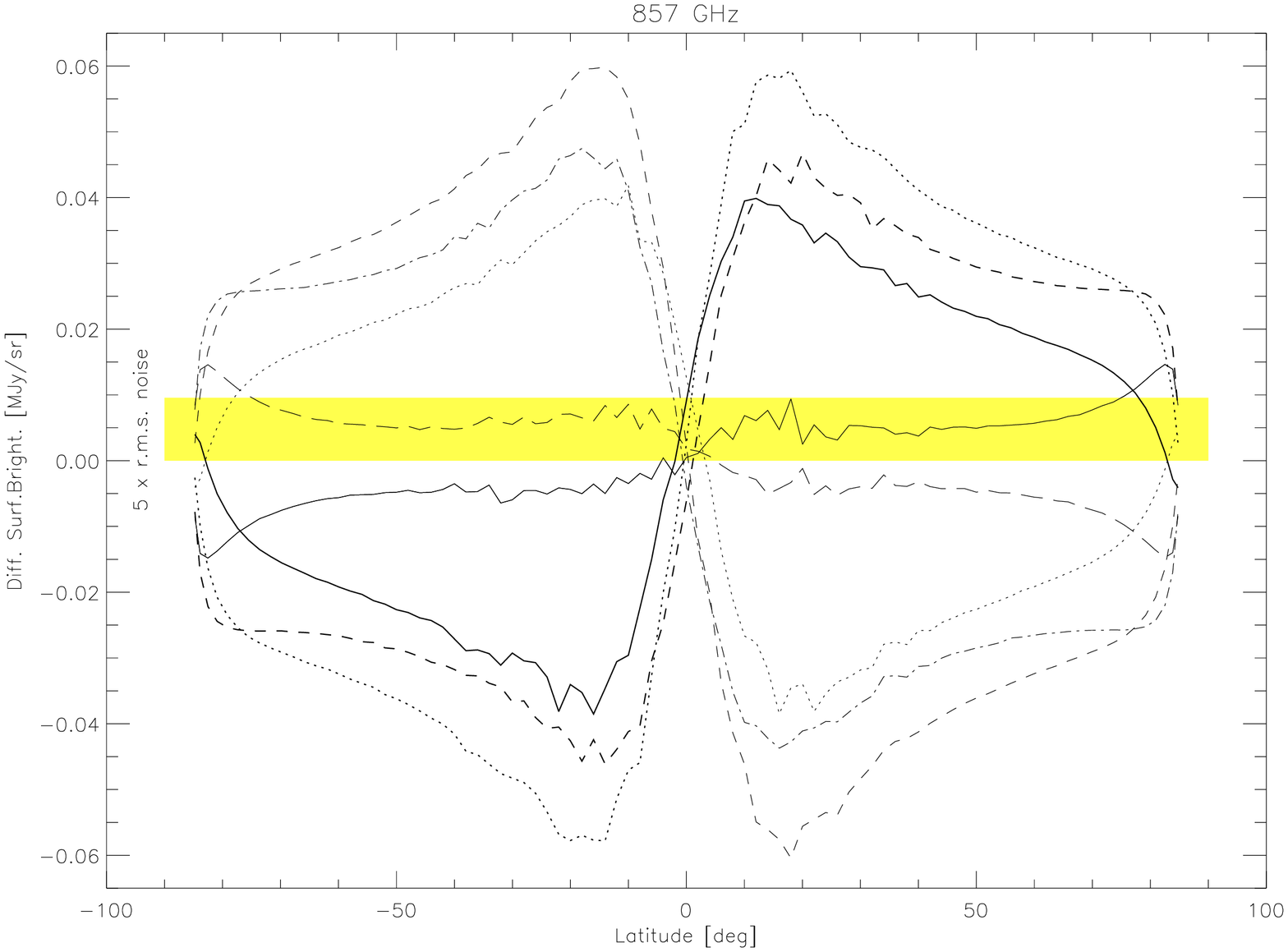}}}
 \caption{
 Differential surface brightness of the ZLE from the Smooth component in 
the 857~GHz channel, $\Ef = 0.65$ 
 calculated for a spacecraft ecliptical longitudes 
        77.7$^\circ$ (solid line),
       122.7$^\circ$ (dotted line),
       167.7$^\circ$ (dashed line),
       212.7$^\circ$ (dash dot line),
       257.7$^\circ$ (long dashed line),
       302.7$^\circ$ (solid thick line),
       347.7$^\circ$ (dotted thick line),
       392.7$^\circ$ (dashed thick line).
 The ascending node of the symmetry plane for the Smooth component cloud
is assumed to be at a longitude of
 77.7$^\circ$, so that these longitudes are equivalent to an angular distance of 
 of 0$^\circ$, 45$^\circ$,  
 90$^\circ$, 135$^\circ$, 180$^\circ$, 225$^\circ$, 270$^\circ$, 315$^\circ$ 
 from the node respectively.
 The surface brightness scale is in units of MJy/sr. 
 The yellow (light-gray in bw) band represents the $5\sigma$ sensitivity (white noise limited)
 expected from {\sc Planck} in the 857~GHz frequency channel for circular patches of $2^\circ$
 of radius. 
\ColorFigureStat
 }\label{fig:zle:diff}
 \end{figure}
} 

\def\FIGCALIBERR{
 \begin{figure}
 \centering
 \resizebox{\hsize}{!}{\includegraphics{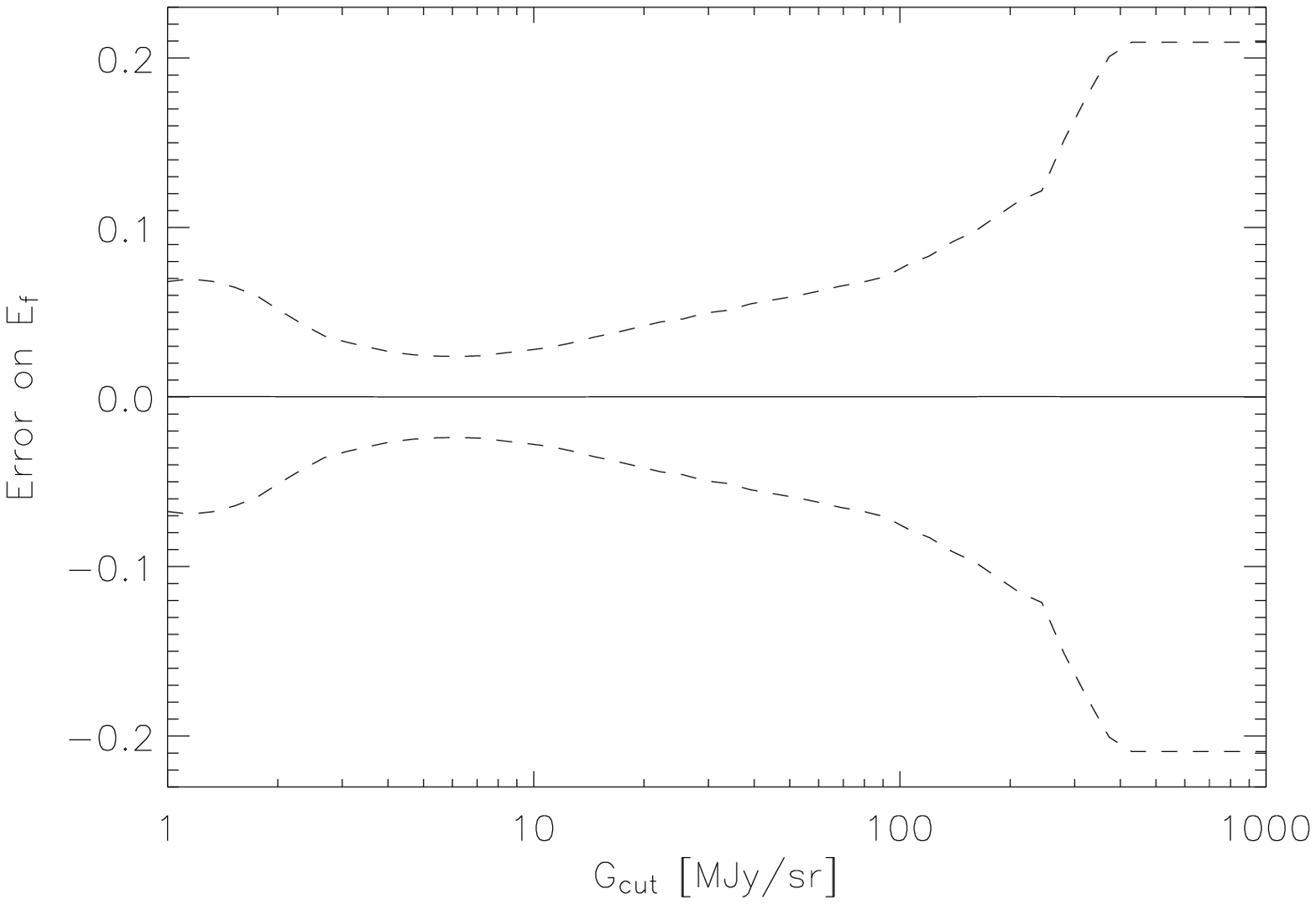}}
 \caption{
Effect of the random relative calibration error on the \Ef\ determination
with the differential method at 857~GHz for 2$^\circ$ patch, as a function
of the cut on the Galactic surface brightness, and assuming a random 
relative 
calibration error with null expectation and RMS per patch of $1\%$. 
The full line represented the expected bias in the \Ef\ determination.
The dashed lines delimit the $1\sigma$ expected variability, both
expectation and RMS are calculated over $3\times10^4$ Monte Carlo 
realizations.
For surface brightnesses $> 10$~MJy/sr the RMS of the 
error decreases decreasing the accepted surface brightness, at lower 
surface brightnesses this is
no longer true due to the reduction in statistics.
 }\label{fig:calib:err}
 \end{figure}
} 

\begin{document}

 \title{
 Zodiacal Light Emission in the {\sc Planck} mission
   }

 \author{M.~Maris\inst{1} 
        \and
        C.~Burigana\inst{2}
        \and
        S.~Fogliani\inst{1}
 }

   \offprints{M. Maris}
   \institute{
 INAF-OATs, 
 Via G.B.~Tiepolo 11, I-34131, Trieste, Italy
      \email{~maris,~fogliani~@oats.inaf.it}
 \and
 INAF-IASF Bologna, 
              via P.~Gobetti, 101, I-40129 Bologna, Italy\\
              \email{~burigana@iasfbo.inaf.it}
             }

   \date{Sent June 28, 2005; revised version January 25, 2006; accepted 
January 27, 2006.}

  \abstract
{The \Planck\ satellite, scheduled for launch in 2007, will produce a
set of all sky maps in nine frequency bands spanning from 30~GHz to
857~GHz, with an unprecedented sensitivity and resolution. 
Planets, minor bodies and diffuse interplanetary dust
will contribute to the (sub)mm sky emission 
observed by \Planck, representing a source of
foreground contamination to be removed before extracting the
cosmological information. }
{The aim of this paper is to assess the expected level of contamination 
in the survey of the forthcoming \Planck\ mission.}
{Starting from existing far-infrared (far-IR) models of the 
Zodiacal
Light Emission (ZLE),
we present a new method to simulate the time-dependent
level of contamination from ZLE at \Planck\ frequencies.}
{We studied the possibility of \Planck\
to detect and separate the ZLE contribution from the other
astrophysical signals.} 
{We discuss the conditions in
which \Planck\ will be able to increase the existing information 
on the ZLE and IDP physical properties.
\\{\it This work is done in the framework of the \Planck/LFI activities.}
}
 %

 %

 %

  \def\sep{--}

   \keywords{
 Interplanetary Medium \sep\
 Infrared: Solar System \sep\
 Submillimeter \sep\
 Methods: numerical \sep\
 Space vehicles: instruments \sep\
 (Cosmology): Cosmic Microwave Background 
   }

 \authorrunning{M.~Maris, C.~Burigana, S.~Fogliani}
 \titlerunning{ZLE in the \Planck\ Mission}

   \maketitle



 \section{Introduction}

The ESA \Planck\ satellite~\footnote{http://www.rssd.esa.int/planck} 
\citep{Tauber:2003}, scheduled for launch in 2007 \footnote{
While this paper was at the end-stage of the editorial process,
   the launch has been posponed to 2008. This will not affect the
   results in this paper.
}, is a full-sky surveyor dedicated to 
cosmic microwave background (CMB) and
millimetric (mm) and sub-mm astronomy. It is  a third 
generation microwave mission,
after the NASA COBE and WMAP missions. 
The surveyor will observe the sky through a 1.5~m
Gregorian aplanatic telescope carrying two
instruments  on the focal surface 
operating at the frequency bands centred at 30, 44, and 70~GHz 
(Low Frequency Instrument, LFI; \citealt{Mandolesi:etal:1998}) 
and 100, 143, 217, 353, 
545, and 857~GHz (High Frequency Instrument, HFI; \citealt{Puget:etal:1998}). 
\Planck\ will be
injected in a Lissajous orbit around the Sun-Earth Lagrangian
point \LT\ of the Sun-Earth system at a distance of $1\,507\,683$~Km~$\sim 
0.01$~AU from the Earth,
from which it will observe the
microwave sky for at least 15~months, necessary to complete two surveys 
of the whole sky with all the receivers. 
The LFI beams are located 
on the \Planck\
telescope field of view 
in a ring with a radius of about 4$^\circ$ 
around the telescope line of sight (LOS)
pointing at a scan angle $\alpha=85^\circ$ from the satellite spin
axis. 
The HFI beams, located closer to the centre, may be
also at few degrees from the LOS.
In the simplest scanning strategy, the spin axis, chosen
pointed in the opposite direction from the Sun,
will be kept parallel to the Sun-spacecraft direction,
re-pointed by $\Delta \theta_s = 2.5'$ once an hour
($1^\circ$ per day) in order to follow the revolution of the \LT\
Lagrangian point, and will spin at 1~RPM.
\Planck\ will scan the sky in nearly great circles approximately
orthogonal to the ecliptic plane at a rate of 24 circles per day, each
circle being scanned consecutively 60 times per hour. In this way
\Planck\ will produce at least two full sky maps for each
frequency channel with an unprecedented resolution (FWHM from
$\simeq 33'$ to $\simeq 5'$) and sensitivity (in the range of
$\simeq 10-49$~mJy on a FWHM$^2$ resolution element).
Although the detailed scanning strategy is currently
under study \citep{Dupac:Tauber:2005}, its general properties imply that 
only objects located at $\sim 80^\circ-90^\circ$ from the Sun will enter 
the large circles traced in the sky by the main beams.
Solar System objects then will be observed nearly in quadrature with the 
Sun i.e. within few degrees from a solar elongation of $\simeq 85^\circ$.

\FIGSCANGEOM

Why would a cosmological mission like \Planck\ consider
the Solar System in its scientific program? How should
Solar System studies take advantage from a mission like
\Planck?
A valuable contribution has been provided by the COBE mission to
solar system studies \citep{Kelsall:etal:1998}. The
scanning strategy of \Planck\ assures that all the Solar
System components located outside the Earth Orbit,
except Mars, will enter at
least once into the field of view of the surveyor during the mission.
In this way point-like Solar System objects 
(external planets;  \citealt{Burigana:etal:2002} , \citealt{Page:etal:2003}, \citealt{Schaefer:etal:2004},
asteroids; \citealt{Cremonese:etal:2002}, \citealt{Schaefer:etal:2004}, comets) and the thermal emission of the
diffuse
interplanetary dust will represent foregrounds which
have to be detected and properly removed in order to avoid the
introduction of systematic errors in the cosmological measures
   \citep{Maris:etal:2003,Maris:etal:2004}.

The {\em Zodiacal Light Emission} (ZLE) due to thermal emission from
the Interplanetary Dust Particles (IDPs) is the far-IR 
counterpart
of the familiar Zodiacal Light due to scattering of the solar
light by IDPs.
Most of the properties of the ZLE below 300~$\mu$m have been
studied by IRAS \citep{Wheelock:etal:1994}, COBE
\citep{Kelsall:etal:1998} and ISO
\citep{Reach:Abergel:Boulanger:1996,Reach:etal:2003}.
Peaking at $\lambda \approx 10$~$\mu$m, the ZLE is one of the
major contributors to the sky background in the far-IR domain at
low ecliptic latitudes.
A first detection in the $300\;\mu\mathrm{m} - 1000\;\mu\mathrm{m}$
band has been assessed 
by \citet{Fixsen:Dwek:2002} using yearly averaged
COBE/FIRAS data.
Even a quick look at the data reported in
\citet{Kelsall:etal:1998} and \citet{Fixsen:Dwek:2002}
allows us to predict a contribution
of ZLE in the 857~GHz channel of \Planck\ of $\approx 0.6$~MJy/sr.
It is evident that
at the \Planck\ lowest frequencies
its contribution is much weaker than 
the Galaxy emission. On the other hand,
at the \Planck\ intermediate and high frequencies
the ZLE is significantly weaker than 
the Galactic emission only at low Galactic latitudes while it is 
comparable to it outside the Galactic plane
(for example, near the poles the Galactic emission
is $\sim 1$~MJy/sr at 857~GHz).
The expected ZLE contribution is larger than the instrumental noise 
at the \Planck\ highest frequencies. 
Therefore, a careful analysis of the ZLE in the \Planck\ data
is required.
Since the ZLE varies over angular scales $\gsim 10^\circ$
it can be properly studied by working at a resolution of
$\simeq 1^\circ - 2^\circ$. At this scale 
the expected $1\sigma$ 
sensitivity per FWHM$^2$ at 857~GHz
at the end of the mission is
$\approx 2 \times 10^{-3}$~MJy/sr 
 \citep{Lamarre:etal:2003}.
The extrapolated background from the Galaxy, representing the main large 
scale background component at this frequency, is $\gsim 1$~MJy/sr. 
Because of the different tilt on the ecliptic of the Galactic plane and 
of the IDP cloud, for most of the scan circles observed by \Planck, 
the sky position of the maximum of the ZLE will fall close to that of the
minimum of the Galactic emission. In this case the ZLE is extrapolated 
to be about half of the Galactic emission.
 Of course one may wonder whether ground-based or balloon born CMB
experiments may have been able to detect such a contribution. 
Looking at some of the most recent balloon-borne experiments, 
ARCHEOPS \citep{Benoit:etal:2003} has constructed maps by bandpassing the 
data between 0.3 and 45~Hz, corresponding to about 30$^\circ$ and $15'$ 
scales, respectively.
MAXIMA \citep{Lee:etal:2001} covered the multipole range 
$36 \leq \ell \leq 1235$ or angular scales smaller that $6^\circ$.
The CMB power spectrum of BOOMERANG
\citep{Netterfield:etal:2002} covers 
multipoles from  $\ell \simeq 75$ to $\ell \simeq 1025$,
equivalent to angular 
scales $10'- 2.4^\circ$. 
From ground based experiments,
DASI 
\citep{Pryke:etal:2002}
measured the power spectrum for 
$100 < \ell < 900$ or angular scales less than $1.8^\circ$.
Among these experiments it seems that only ARCHEOPS is able to 
detect the large scale brightness variation connected with the ZLE.
We then expect that the ZLE will be observable as an excess of
signal superimposed on the Galaxy or it may be 
considered as a source of systematics in
studying the large scale Galactic emission.

Our strategy, in line with past studies, exploits the existing far-IR 
observations, 
included in models, to derive the spatial distribution of the ZLE and to 
extrapolate 
its Spectral Energy Distribution (SED) at \Planck\ frequencies.
We take as a reference the yearly averaged values of 
 \citet{Fixsen:Dwek:2002}.
Our starting point to model the spatial distribution of the ZLE, 
the work of \citet{Kelsall:etal:1998} for the ZLE  based on 
the COBE data (hereafter indicated as the {\em COBE-model}), has many 
similarities with the 
IRAS model by \citet{Wheelock:etal:1994}.
It describes in detail the emissivity of the IDP cloud, assumed to 
extend up to 
$\simeq 5.2$~AU from the Sun, for wavelengths up to about 
300~$\mu$m. According to the COBE-model four components contribute to 
the ZLE: 
 the dominating smooth component,
 the Earth orbit locked ring of dust (or circumsolar ring),
 the trailing blob,
 and three bands of dust.

In this work only the standard IDP component has been considered. The analysis of the 
plausible, but not yet determined, contribution from the Kuiper Belt dust grains
\citep{Landgraf:etal:2002} will be the subject of another work.

\FIGSCANGEOMB

With respect to other foregrounds usually considered in CMB studies,
the ZLE (as the other Solar System objects) is peculiar, depending for
its surface brightness not only on the pointing direction but also on
the instantaneous position of the observer within the Solar
System.
Galactic and extragalactic foregrounds are generated by
sources located so far from the observer that
parallactic effects due to the motion of the observer within the
Solar System are negligible compared to the
instrumental resolution of CMB observatories. 
On the contrary, since the observer is located within the Solar
System, the orbital motion about the Sun leads to changes in the
ZLE brightness distribution as a function of the pointing
direction.
This underlines the relevance of studying the ZLE not only on maps but
also on time ordered data streams (TODs). Moreover, the accurate 
simulation of
observations for a satellite mission like \Planck\ cannot be
based on maps since the details of the orbit will have to be
considered in addition to the usual scanning law.

The main aim of this work is to contribute to the following subjects:
\first\ to define a representation method for the ZLE suitable
for map based CMB mission simulators, with particular relevance for
the \Planck\ mission;
\second\ to determine suitable approximations (like scaling frequency
laws) for the simulation of this component in the framework of the
\Planck\ simulation pipeline;
\third\ to determine to what extent the ZLE will impact the 
\Planck\ survey;
\fourth\ to determine if it will be possible to separate the
contribution of the ZLE from the data produced by the \Planck\
mission in a self-consistent manner (i.e. reducing as much as 
possible the introduction of priors based on results from other missions 
in the \Planck\ data processing pipeline);
\fifth\ to explore the possibility of \Planck\ to produce useful
scientific results about the ZLE at frequencies barely explored in the past.

The paper is organised as follows. In Sect.~\ref{sec:theory} we briefly 
review the present knowledge about the ZLE. 
Sect.~\ref{sec:model} describes the framework of our simulations, mainly based on the model of \citet{Kelsall:etal:1998}, 
and the details of our numerical code discussing its main assumptions 
in the light of recent theoretical results. 
In Sect.~\ref{sec:zle:serie} we present a series expansion of the ZLE 
spatial dependence that can be useful for many simulations and data 
analysis applications. 
In Sect.~\ref{sec:results} we 
describe the main results of these simulations (mainly in form of TODs 
and maps) and 
compare the ZLE contribution to those expected from 
the Galactic emission. 
Sect.~\ref{sec:separability} is devoted
to the separation of ZLE in the \Planck\ data. Particular care is 
given to the analysis of the systematic effects in the differential
approach for ZLE separation.
Our main results and conclusions are summarised in 
Sect.~\ref{sec:conclusions}.

 \section{Physical and geometrical properties of ZLE}\label{sec:theory}

To assess the expected errors in predicting the ZLE surface brightness at 
\Planck\ 
frequencies from a model based on far-IR and IR data
we review some theoretical concepts needed to link 
the ZLE model, and in particular the COBE-model, to the optical properties
and the size distribution and other physical properties of the IDPs. 
In sub-mm and mm bands the dominant emission mechanism from IDPs is thermal 
emission of IR radiation driven by solar heating. 
The most general expression for the brightness 
averaged over the bandwidth and the beam, 
detected in a radiometric channel of frequency $f$ 
on-board a space-born experiment produced 
by a given population of IDP grains, representing a component $c$\ of the
IDPs cloud is

 \begin{equation}\label{eq:b:integral:full}
 \begin{array}{l}
  \lefteqn{I_{f,c}(\Point,\SpaceCraft) = 
   \left[
      \int_{0}^{+\infty}   \int_{\Omega}  d\nu \,
        d\Point' \, W_f(\nu )\, G_{f}(\Point',\nu)
   \right]^{-1}
} \\  
 \\
 \cdot \,
    \int_{0}^{+\infty} 
    \int_{\Omega} d\nu \, d\Point' \, 
       \Bigl[
     W_f(\nu ) G_{f}(\Point' - \Point,\nu) \,
 \\
 \\
 \cdot \,
 \int_{0}^{+\infty}   
\int_{0}^{+\infty} da \, ds \\
 \left. 
 \cdot \,
\left( B_{\nu}(T_c(\mathrm{r}(s))
  \pi a^2 \Qabs_c(\nu,a,\mathbf{r}(s)) \,
                           \frac{d\,N_c\left(a,\mathbf{r}(s)\right)}{da} \right) \right] \, ,
 \\
\end{array}
\end{equation}

\noindent
 here
 $a$ is the grain size, $\nu$ the frequency, 
 \Point\ is an observing direction, 
 $s$ the
 distance from the observer along \Point, 
 \SpaceCraft\ the position 
 of the observer respect to the Sun,
 $\mathbf{r}(s) = s\Point + \SpaceCraft $ the position within the Solar System respect to  
 the Sun along the line of sight, 
 $W(\nu)$ is the instrumental frequency response,
 $G_\nu(\Point)$ the beam response, 
 $T_c(\mathbf{r})$ is the thermodynamical temperature for grains in the population $c$,
 $B_\nu$ the blackbody emissivity and
 $\Qabs_c(\nu,a,\mathbf{r})$ is the absorption coefficient for the population $c$ of 
 grains,
  ${d\,N_c(a,{\mathrm{r}})}/{da}$ the size distribution for grains.
A rigorous calculation of this integral is difficult since it 
requires the knowledge of many poorly constrained quantities.
Alternatively effective models, such as the COBE model,
are used in place of Eq.~(\ref{eq:b:integral:full}). 
We discuss here the relation between the two approaches.
Let us assume that:
 \first\ the size distribution and the grain optical properties do not depend 
 on $\mathbf{r}$, $\Qabs_c(\nu,a,\mathbf{r}) \approx
\Qabs_c(\nu,a)$, ${d\,N_c(a,\mathbf{r})}/{da} \approx dn_c(a)/da
\cdot N_c(\mathbf{r})$ with $N_c(\mathbf{r})$ the spatial distribution of grains;
 \second\ the beam is symmetrical and it does not depend on $\nu$ within the frequency bandwidth
 of each channel $f$;
 \third\ the beam is small compared to the typical angular scales over which
the ZLE varies.
With these assumptions Eq.~(\ref{eq:b:integral:full}) simplifies to

\begin{equation}\label{eq:b:integrall:simple}
 \begin{array}{ll}
  I_{f,c}(\Point,\SpaceCraft) = &
                    K_{f}\, 
                    E_{f,c} \,
                    n_{0,c} 
             \\ &\\ & 
                    \cdot \int_{0}^{+\infty} ds
                        \, B_{f}(T_c(\mathbf{r}(s))
                        \, N_c(\mathbf{r}(s)),
\end{array}
\end{equation}

 \noindent
where $\Kf$\ is a {\em color correction} which takes into
account the averaging over the frequency weighted by the 
instrumental response (see Sect.~\ref{sub:sec:Kf}),,
$n_{0,c}$\ is the optical density 
for the given dust component, $E_{f,c}$ is an 
{\em emissivity correction} related to the size distribution and
composition of grains.
For multifrequency observations, \Ef\ is usually 
normalised to a reference frequency, $f_0$.
For example in \citet{Kelsall:etal:1998} \Ef\ is normalised to the value it has in
the COBE/DIRBE 25~$\mu$m ($f_0=12\,000$~GHz) channel. 
Note that such kind of normalisation of \Ef\ implies a corresponding 
renormalisation of
the optical density 
$n_{0,c}$ in Eq.~(\ref{eq:b:integrall:simple}).
In principle it would be possible to compute $\Qabs(\nu,a)$\ and then \Ef\
from Mie theory assuming appropriate grain shapes and composition 
and using commonly available software \citep{Bohren:Huffman:1998},
as done recently in the study by \citet{Reach:etal:2003} of the 
ZLE emissivity at $\lambda < 100$~$\mu$m.
However, the composition and shape of IDP grains 
in the size range relevant at \Planck\ frequencies are only poorly 
known. 
We prefer to determine \Ef\ at \Planck\ frequencies comparing
theoretical estimates with the existing COBE/FIRAS data,
as detailed in the next section.

 \TABLEEFFD

 \subsection{Estimating $E_f$ at \Planck\ frequencies}\label{subsec:cobe:Ef}

We estimated \Ef\ in the relevant range of frequencies comparing
the expected ZLE yearly averaged from the COBE model 
with the existing surface brightness measures at \Planck\ frequencies 
from \citet{Fixsen:Dwek:2002} based on COBE/DIRBE and COBE/FIRAS 
data.
However, given the uncertainties in the interpretation of 
these data,
other extrapolation methods are possible.
According to the main result of \citet{Fixsen:Dwek:2002}, 
the SED of the ZLE Smooth component is approximately similar to 
a blackbody with $T \approx 240$~K scaled by an emissivity factor
nearly constant for $\lambda < 160$~$\mu$m and scaling
as $\lambda^{-2}$ at longer wavelengths: 

 \begin{equation}\label{eq:FD:emissivity}
 \begin{array}{ll}
 \langle I_{\nu=f} \rangle_{\mathrm{year}}^{\mathrm{FD}} \approx  &  3 \times 10^{-7} \times B_\nu(T=240\;\mathrm{K}) \\ &\\
  & 
 \times
  \left\{
   \begin{array}{ll}
    1, & \nu > 1875 \;\mathrm{GHz}, \\
    \left(\frac{\nu}{1875\;\mathrm{GHz}}\right)^2, & \nu \le 1875 \;\mathrm{GHz}\\
 	\end{array}
 \right\} \,          .
 \end{array}
 \end{equation}

 \noindent
As evident from this analysis of COBE/FIRAS data \citep{Fixsen:Dwek:2002}, 
the accuracy of the recovered ZLE spectrum is good at frequencies higher
than about 800~GHz while the error bars
significantly increase at lower frequencies, where most of the \Planck\ frequency
channels are located.
This leaves space for further improvement in the field with \Planck.
After estimating the yearly average surface brightness from the COBE model,
$\langle Z_{f} \rangle_{\mathrm{year}}$, assuming 
$\Ef = 1$ we obtain

 \begin{equation}\label{eq:Ef:FD}
  E^{\mathrm{FD}}_{f, \;\mathrm{Smooth}} = \frac{\langle I_{\nu=f} \rangle_{\mathrm{year}}^{\mathrm{FD}} }
        {\langle Z_{f} \rangle_{\mathrm{year}}} .
 \end{equation}

 \noindent
The resulting estimated \Ef, normalized to 1 at $f=1.2\times10^4$~GHz,
are reported in Table~\ref{tab:Ef:FD} 
with the expected yearly average surface brightnesses. 
$E^{\mathrm{FD}}_{f, \;\mathrm{Smooth}}$ is 
approximated in the range of frequencies of interest by 

 \begin{equation}\label{eq:Ef:FD:Fit}
  E^{\mathrm{FD}}_{f, \;\mathrm{Smooth}} \simeq 126.63 \left(\frac{f}{1.2\times10^4\mathrm{GHz}}\right)^2 \;.
 \end{equation}

 \subsection{The colour correction}\label{sub:sec:Kf}

The colour correction \Kf\ in Eq.~(\ref{eq:b:integrall:simple}) takes 
into account 
the effect of the frequency instrumental response 
within the bandwidth.
At \Planck\ frequencies, most of the ZLE is due to IDPs with
temperatures exceeding $\approx 200$~K; within the bandwidth 
of each \Planck\ frequency channel, their emission 
integrated along the line of sight can be approximated by a power law with 
spectral index $\gamma_f$ and normalisation $\densFf(\nu=f) = F_f$
(from here we will omit the
pointing dependence to simplify the notation).
In addition, within a frequency band 
$\densFf(\nu) = \epsilon_f(\nu) \mathcal{Z}_f(\nu)$,
where 
$\epsilon_f(\nu) = (\nu/f)^{\gamma_{f,\epsilon}} E_f$
is the {\em spectral emissivity correction}
and
$\mathcal{Z}_f(\nu) = (\nu/f)^{\gamma_{f,z}} Z_f$
accounts for the spatial distribution expected from grains emitting
as blackbodies and can be separated into a pure spatial dependence $Z_f(\nu)$
and a pure frequency scaling.
Here normalisations are defined to match 
the values of $E_f$ and $Z_f$ for $\nu = f$ and 
the overall $\gamma_f$ turns to be
$\gamma_f = \gamma_{f,\epsilon} + \gamma_{f,z}$.
Therefore, by imposing

 \begin{equation}\label{eq:brightness:integral:approx}
  F_f \Ef \Kf \approx 
  \frac{1}{\Delta f}
  \int_{\mathrm{BW}} 
    W_f(\nu) \epsilon(\nu) \mathcal{F}_f(\nu) d \nu 
 \end{equation}

 \noindent
and Taylor expanding in the 
$\log$~-~$\log$ space about $\log f$ the brightness
and the emissivity correction, we obtain 

 \begin{equation}\label{eq:brightness:integral:approx:expanded}
  \Kf  = 
  \frac{1}{\Delta f}
          \int_{\mathrm{BW}} W_f(\nu) 
           \left( 
             \frac{\nu}{f}
            \right)
 ^{\gamma_{f,\epsilon} + \gamma_{f,z}}
            d \nu 
            \; , 
 \end{equation}

 \noindent
together with 
  $F_f = \mathcal{F}_f(f)$ and 
  $\Ef  =  \epsilon(f)$.
The validity of the approximation represented by
Eq.~(\ref{eq:b:integrall:simple}) is then verified.

We assess the value of \Kf\ needed to compensate
the fact that in Eq.~(\ref{eq:b:integrall:simple}) the integration
over the bandwidth in Eq.~(\ref{eq:b:integral:full}) has been 
neglected. 
In addition we want to assess the level of uncertainty in the \Kf\ 
correction induced  by the uncertainty in the \Ef\ prediction.
We restrict ourselves to the 
illustrative case of a simple top-hat window $\Wf(\nu)$ with relative bandwidth 
$r_f = \Delta f / f$. 
For \Planck\ $r_f = 0.2$ for $f\le70$~GHz and $r_f = 0.25$ for 
$f\ge100$~GHz. 
We tested that for different reasonable shapes of $\Wf(\nu)$ the results do not change
significantly. 
Under these conditions, for $\gamma_f \neq 1$
 \begin{equation}\label{eq:Kf:square}
 \Kf = \frac{\WfZ}{r_f (1+\gamma_f)}
      \left[
      \left(
      1 + \frac{r_f}{2}
      \right)^{1+\gamma_f}
      -
      \left(
      1 - \frac{r_f}{2}
      \right)^{1+\gamma_f}
      \right],
 \end{equation}

 \noindent
where $\WfZ$ is the normalisation for the spectral window.
 
For numerical estimates, assuming $\WfZ = 1$ and 
$\gamma_{f,z} \approx \gamma_{f,\epsilon} \approx 2$, 
we obtain 
 $K_{f \le 70\;\mathrm{GHz}} = 1.020$ and $K_{f \ge 100\;\mathrm{GHz}} = 1.031$.
For the case of a frequency-independent \Ef, $\gamma_{f,\epsilon} = 0$ and
we find $K_{f \le 70\;\mathrm{GHz}} = 1.003$ and $K_{f \ge 100\;\mathrm{GHz}} = 1.005$.
Leaving $\gamma_f$ to vary within $\pm1$ unit \Kf\ changes by only $2\%$.
Thus it will be possible to avoid applying this small
correction in the numerical estimates presented in the remaining part of this paper. 

Note that in \citet{Kelsall:etal:1998} the colour correction
is defined as a correction for the instrumental response when a blackbody
is observed. 
For this reason the colour correction is parametrised
as a function of the blackbody temperature $T$ which is a function of
the position along the line of sight, so that the argument of the
pointing direction
integral in Eq.~(\ref{eq:b:integrall:simple}) would have to be 
scaled by the spatial dependent $\Kf(T)$.
However, at our frequencies the bulk of the blackbody emissivity 
comes from grains emitting not too far from the Rayleigh-Jeans
limit, i.e. with a frequency power law scaling 
(within our rather limited bandwidth) with a power law index largely 
independent of $T$. 
In this case the two definitions for \Kf\ are equivalent.

 \section{The model and the numerical code}\label{sec:model}

 \FIGSMOOTHCNT

The surface brightness calculated for a given frequency band is 

 \begin{equation}\label{eq:zle:decomposition}
 \begin{array}{ll}
    F_{ZLE,f}(\Point,\Sun,\SpaceCraft) &= \sum_c F_{c,f}(\Point,\Sun,\SpaceCraft) \\
&\\
   &=
    \sum_c E_{c,f} Z_{c,f}(\Point,\Sun,\SpaceCraft) \, .
 \end{array}
 \end{equation}

 \noindent
Here $\Sun$ denotes the position of the Sun within 
the Solar System and the index $c$ denotes 
the specific component either 
 the dominating smooth component,
 the Earth orbit locked ring of dust,
 the trailing blob,
 or one of the three bands of dust.
According to the discussion in Sect.~\ref{sec:theory}, 
in Eq.~(\ref{eq:zle:decomposition}) 
we separate the calculation 
of the spatial distribution of the ZLE assumed to be a blackbody 
from the more uncertain emissivity correction. 
The detailed COBE MODEL has been already described in the literature
 \citep{Kelsall:etal:1998,Fixsen:Dwek:2002} 
and only few details need to be reviewed.
The relevant geometry of ZLE observations is shown in Fig.~\ref{fig:scanning:geometry}: 
the Sun (or the barycentre of the Solar System \SSBar), 
the position of the \LT\ point, the spacecraft 
(in this case \Planck)
position \SpaceCraft, the centre of the distribution of
dust related to the component $c$ of interest, \ComponentCenter,
and the related vectors drawn between these points. 
The most important among them are:
  the position of a point at distance $s$ from
the spacecraft along the pointing direction \Point: 
 $\mathbf{R}(s) = \SpaceCraft + s \Point$; 
its position with respect to the Sun, 
 $\mathbf{R}_{\mathrm{s}} = \mathbf{R} - \Sun$, 
and to the centre of the cloud defined by 
 $\mathbf{R}_{0,c}$: $\mathbf{R}_c' = \mathbf{R} - \mathbf{R}_{0,c}$. 
In addition we define 
 $R_c' = |\mathbf{R}_c'|$. 
 $X_c'$, $Y_c'$, $Z_c'$ 
denote the Cartesian components of $\mathbf{R}_c'$ and 
$X$, $Y$, $Z$ those of $\mathbf{R}$. 
The model specifies the 3D dust density distribution of each component, 
factorised into a radial dependence and a vertical dependence, 
as the heliocentric dependence of the mean dust temperature. 
In this way, for each component $c$ it is possible to define a proper reference frame 
about which the cloud $c$ has cylindrical symmetry. Its origin
coincides with the centre of the cloud
and its midplane with the equatorial plane of the cloud; it may be 
rotated, tilted and shifted with respect to the ecliptic plane.

We have implemented the COBE MODEL 
in a {\tt FORTRAN-90/95} program called \FSZOD\
({\sc Flight Simulator - Zodiacal Light 
Emission})
embedded in a supporting \OCTAVE\ pipeline
 \citep{Report:FsZod}.
The code was originally designed as a module of the \Planck\ 
Flight Simulator but can be used for any other experiment.
Since the \Ef\ scalings are largely uncertain it is left to the
user to apply the proper one to the output of these programs.
In this work we use the code to study the time dependence in the
signal acquired by \Planck\ for the nominal scanning strategy 
\citep{Dupac:Tauber:2005} 
 and a recently simulated spacecraft orbit \citep{Orbita:2002}
 and to predict the ZLE
induced perturbation in \Planck\ data. 
We choose to express our results in terms of brightness (MJy/sr), 
as for IRAS and COBE data.

 \subsection{Series expansion of the ZLE spatial dependence}\label{sec:zle:serie}

\FIGONE

 Simulators for CMB missions and the related data-reduction
pipelines, such as those realized for \Planck, are largely based on
maps. 
A map allows a good representation of the sky brightness as
a function of the pointing direction. 
This procedure neglects the time-dependent information on the 
\Planck\ position within the Solar System, leading to a loss of
information when Solar System components are considered.
The ground segment of a mission like \Planck\ would be 
able to handle and analyse TODs
as well as maps obtained from them 
\citep{Pasian:Sygnet:2002,CMB:On:Rings:I,CMB:On:Rings:II}. 
A TOD of ZLE would allow an exact representation of any
seasonal dependence.
However, TODs are large and their realization requires an 
effective scanning strategy and satellite orbit, possibly accommodated 
during the mission, so that the exchange of simulated data in the form of
TODs is not practical for the data analysis of a multichannel, 
high-resolution mission like \Planck. 
We implement a method able to
 1.~properly represent seasonal effects in a large set of mission configurations; 
 2.~possibly be applied to other missions; 
 3.~exploit the (cylindrical) symmetries in the components of the IDP cloud.

We propose to generalise the concept of pixelized map.
A pixelized map is usually defined as the values assumed by a given observable
on a set of pixels ordered according to the adopted pixelization scheme. 
In this contest we can introduce a generalisation of this concept by 
defining a pixelized \map\ as a list of values assumed by a given 
observable on a set of pixels which are also functions of the positions of 
the Sun $(\Sun)$ and of the Spacecraft (\Planck\ in our case)
$(\SpaceCraft)$ within the Solar System.
In a reference frame (r.f.) in which 
the displacements of the Sun and spacecraft positions 
are just small fractions of their averaged positions, 
the pixelized \map\ can be replaced by the series expansion of 
the observable about the average positions of the Sun and the
Spacecraft.

Then, denoting with \ASpaceCraft, \ASun\ the reference positions of
Spacecraft and Sun about which the series expansion is performed,
we have 

 \begin{equation}
      \SpaceCraft  =  \ASpaceCraft + \DSpaceCraft,
      \;\;\;\;
      \Sun         =  \ASun + \DSun.
 \end{equation}

 \noindent
In the numerical computations for \Planck\ we exploit values of the displacements along each
direction up to about $\pm 0.03$~AU for \DSun\ and up to about $\pm 0.07$~AU for \DSpaceCraft,
about a factor of two 
wider than any reasonable displacement for a mission that will reside near \LT.
In addition, we adopt in this work the {\tt HEALPix} scheme
\citep{Gorsky:etal:2005} with ``ring'' ordering,
widely used in  the CMB community. 

We have tested that in the case of the ZLE it is preferable to expand 
not the brightness spatial distribution but its logarithm in power series.
Then, denoting by $Z_{f,c,p}(\Sun,\SpaceCraft)$ the brightness
integral for a given frequency channel $f$, component $c$ and
pixel index $p$ (connected to the pointing direction $\Point$ by the
mapping scheme) as a function of \Sun\ and \SpaceCraft\ we adopt
the following decomposition

 \begin{equation}
 \begin{array}{lcl}
   Z_{f,c,p}(\Sun,\SpaceCraft)  & = & 
   \overline{Z}_{f,c,p}
   F_{\mathrm{S},f,c,p}(\DSun) 
F_{\mathrm{P},f,c,p}(\DSpaceCraft)
 \\
 & & \\
 & & \cdot    
   F_{\mathrm{PS},f,c,p}(\DSun,\DSpaceCraft),
 \end{array}
 \end{equation}

 \FIGYEARLYAVER

 \noindent
where   $\overline{Z}_{f,c,p} \equiv
Z_{f,c,p}(\ASun,\ASpaceCraft)$ and
$F_{\mathrm{S},f,c,p}(\DSun)$,
$F_{\mathrm{P},f,c,p}(\DSpaceCraft)$,
$F_{\mathrm{PS},f,c,p}(\DSun,\DSpaceCraft)$ are exponential
functions of polynomials of \DSun\ and \DSpaceCraft.
In order to achieve an accuracy better than $\simeq 1\%$ 
for more than 96\% of the pixels, 
we verified that~\footnote{The remaining pixels will have an accuracy 
worse than 1\% (the worst accuracy recorded being 2.5\%) in the case of 
a displacement of the spacecraft and of the Sun at the limits of the region for which
the expansion is calculated.}
it is sufficient to consider the following terms of the series expansion:

 \begin{eqnarray}
     \log F_{\mathrm{S},p}(\DSun) &=&
       a_{\mathrm{S},p,i} \DSuni \\ 
 &&\nonumber+ b_{\mathrm{S},p,ij} \DSuni \DSunj
    \, , \\
    \log F_{\mathrm{P},p}(\DSpaceCraft) &=&
       a_{\mathrm{P},p,i} \DSpaceCrafti \\
       \nonumber
       && + b_{\mathrm{P},p,ij} \DSpaceCrafti \DSpaceCraftj \\
       && + c_{\mathrm{P},p,ijk} \DSpaceCrafti \DSpaceCraftj \DSpaceCraftk \nonumber
    \, , \\
    \log F_{\mathrm{SP},p}(\DSun \DSpaceCraft) &=&
       b_{\mathrm{SP},p,ij} \DSuni \DSpaceCraftj
    \, ,
 \end{eqnarray}

 \noindent
where 
\DSuni\ and
\DSpaceCrafti, $i=X, Y, Z$ are the Cartesian components of \DSun\ and
\DSpaceCraft;
the $a$, $b$, and $c$ terms denote the first, second and third
order coefficients, respectively, and repeated indices $i$,
$j$, $k$ = $X$, $Y$, $Z$ are summed. 
The number of independent coefficients is 3 for
$a_{\mathrm{S},p,i}$ and $a_{\mathrm{P},p,i}$, 6 for
$b_{\mathrm{S},p,ij}$ and $b_{\mathrm{S},p,ij}$, 9 for
$c_{\mathrm{P},p,ijk}$ and $b_{\mathrm{SP},p,ij}$. The other
coefficients are simply obtained by index permutations. 
Then, for each pointing direction a total of 38 independent 
components has to be computed.
These coefficients can be determined by solving 
by least squares a sufficiently large set of 
independent equations obtained by exploiting different combinations of 
displacements $\DSpaceCraft$, $\DSun$ (in the case of \Planck\ about
100).
Metrics are then applied to assess the quality of the data
generated with this series expansion compared to the data generated
by the full simulation 
 \citep[for further details see][]{Maris:etal:2005b}. 
A dedicated support {\tt IDL} library to handle the files 
of expansion coefficients, named {\tt ZLE\_IDL}, has been created
 \citep{Maris:2005}.

 \FIGALONGACROSS

 \section{Results}\label{sec:results}

The contour plot in Fig.~\ref{Fig:Contour:Smooth} is an example
of $Z_f(\Point,\Sun,\SpaceCraft)$ for the Smooth component contribution
calculated for the 857~GHz frequency channel assuming 
$\Ef = 1$. 
According to the discussion in Sect.~\ref{sec:theory},
 $\Ef \approx 0.65$
and the expected observed surface brightnesses are about 
2/3 the values reported in the plot.
The figure represents the variation of $Z_f$ as a function of
the pointing direction
\Point\ for a given combination of \Sun\ ad \SpaceCraft. Contours
are drawn for $Z_f = 0.21$, 0.22, 0.23, 0.24 0.25, 0.30, $\dots$,
3.2~MJy/sr.
Given the cylindrical symmetry of the IDPs, the pointing in the plot is
expressed as a function of the
ecliptical latitude and of the relative ecliptical longitude,
i.e. the difference between the longitude of the pointing
direction and the longitude of the solar direction which is at
the centre of the plot.
The blue dotted line represents the path described by a 
\Planck\ beam at the centre of the field of view assuming the
nominal scanning strategy
and the grey band represents the region observed by considering all the 
\Planck\ beams.
Having the IDPs cloud a cylindrical symmetry, and the symmetry 
reference frame being nearly equivalent to the \Planck\ comoving reference 
frame, for the nominal scanning strategy both the signal contour levels 
and the region observed by \Planck\ will shift approximately in 
the same way when the spin axis is repointed. 
Consequently, only a small fraction of the possible pointings in the 
cloud reference frame will be observed by \Planck.
In the case of more complicated scanning strategies, such as those including 
slow spin axis precession about the Sun - Satellite direction or slow 
oscillations above / below the ecliptic \citep{Dupac:Tauber:2005}, 
the scanning path will be shifted normally and along the ecliptic plane. 
Slow spin axis precession or oscillations with semi-amplitudes of 
$\simeq 5^\circ - 10^\circ$, such as those considered 
for the \Planck\ scanning strategy, will change the path reported in the
plot by $\simeq 5^\circ - 10^\circ$ with a resulting signal difference
more sensitive near the ecliptic plane ($\sim 0.05 - 0.1$~MJy/sr).
However for any reasonable scanning strategy the envelope of all the possible scanning path 
will be only twice or three times wider than the grey band in the figure. 

The tilt of the cloud with respect to the ecliptic plane and parallactic
effects induced by the motion of the spacecraft with respect to the cloud
introduce small modifications in the pattern of the contour lines and
between the signal TODs from different scan circles.
 %

\TABEFUNCONEDEG

TODs may be generated at any desired sampling rate, for example from that corresponding 
to 1/3 of the instrumental FWHM resolution to $1^\circ - 2^\circ$ resolution.
It is important to estimate the error in the computation of the signal in the TODs 
when the true convolution with the beam about its centre direction is replaced by 
the convolution with a ``pencil beam''.
Denoting with $\delta_\perp$ and $\delta_\parallel$ the
displacements from the beam centre respectively
along the direction parallel to the scan circle oriented 
in the scan direction and normal to it towards the Sun,
we compute the $Z_f$ derivatives along these directions.
They are displayed
in Fig.~\ref{fig:along:across}, where $d\ln Z_f/d\,\delta_\perp$ and 
$d\ln Z_f/d \,\delta_\parallel$ are plotted for different positions along 
the scan circle of Fig.~\ref{Fig:Contour:Smooth} at 857~GHz and for the Smooth component.
For the Smooth component 
 $\left|d\ln Z_f/d\,\delta_\perp \right| < 5.6 \times 10^{-3}$~deg$^{-1}$ 
 and 
 $\left|d\ln Z_f/d \,\delta_\parallel \right| < 8.8 \times 10^{-3}$~deg$^{-1}$
(here $Z_f$ is in MJy/sr).
For displacements less than $\mathrm{FWHM}/2 \simeq 2.5$~arcmin the error introduced by this
approximation is at most 0.03\% along $\delta_\perp$ and 
0.05\% along $\delta_\parallel$. 
These error estimates can be linearly rescaled to larger displacements 
$\delta_\perp$ and $\delta_\parallel$ such as those associated with the sky pixelization.

Fig.~\ref{Fig1} represents a portion of a TOD, with the associated uncertainty, 
simulated at 857~GHz without noise and with a sampling at a resolution of 1$^\circ$.
It is the generated sampling
$Z_{\mathrm{Smooth},\;857\;\mathrm{GHz}}$\ from 
Fig.~\ref{Fig:Contour:Smooth} scaled with
$E_{\mathrm{Smooth},\;857\;\mathrm{GHz}} = 0.65$.
In the figure we report for comparison our preliminary estimate of
the sum of the other ZLE components.
Comparing the TOD with the corresponding contour map one sees 
that maxima in ZLE (red lines in the figure) occur when the beam
crosses the plane of the IDPs cloud, slightly below the ecliptic plane
for the considered case. 
Minima instead occur when  the beam is approximately orthogonal to it.
For \Planck, the spin induced modulation of the ZLE signal has a main 
period equivalent to 2 cycles per minute (for the nominal spin rate of
1~r.p.m.).
In a single scan circle, two maxima occur when the beam
crosses the ascending node and the descending node
between the scan plane and the IDPs cloud plane.
Since the plane is tilted on the ecliptic, 
Fig.~\ref{Fig:Contour:Smooth}
allows us to predict the occurrence of asymmetries between the two
peaks even if the scan circle is centred at the antisolar direction.
The secondary components contribute $\approx~10\%$ to the bulk of the ZLE.

To compare the expected ZLE signal with the Galactic foreground, 
TODs for the Galactic emission have been generated and averaged within a 
circle of $1^\circ$ radius. 
Galactic emission TODs are generated using Galactic maps obtained by 
\citet{Schlegel:etal:1998} 
with the prescriptions in \citet{Finkbeiner:etal:1999} for the 
scaling of the Galactic surface brightness as a function of frequency and 
pointing 
direction. 
Due to the tilt of the ZLE symmetry plane over the Galactic plane,
the ZLE at 857~GHz is comparable to the Galactic emission at low
ecliptic latitudes where the Galaxy is weak, contributing a
peak surface brightness of $\approx 0.7_{-0.2}^{+0.4}$~MJy/sr, or 
approximately half of the weakest Galactic signal along that circle.
Of course, the ecliptical longitude of the spin axis about which the scan circle
is drawn will affect the relative contribution of the ZLE with respect to 
the Galaxy. 
For this reason, we report in Fig.~\ref{fig:yearly:aver} 
the variation of the ratio 
$\Ef Z_f/G_f$ (where $G_f$ denotes the Galactic surface brightness -- we 
consider here 
only the dominant dust emission)
with the ecliptical longitude of the spin axis,
for three different \Planck\ frequencies
and for circular patches of $1^\circ$ radius.
The white-full line represents the ratio averaged over the given scan circle.
The black band is the $\pm 3\sigma$ range of such variation, while the highest
ratios expected for each scan circle are represented by the gray-dashed line.
At 857~GHz for about half of the scan circles the 
expected peak ZLE is roughly half of the Galactic dust emission. 
Since the ZLE frequency scaling is not much different from that 
of the Galactic dust 
emission, also at the lower 
frequencies considered here its peak contribution to the sky emission is 
still larger than 
some ten percent 
of the Galaxy for most of the circles. 
This contribution is compared to the instrumental sensitivity. 
The gray band at the bottom of each frame in the figure 
represents the ratio between the
instrumental noise, $N_f$, and the Galactic dust 
$N_f/G_f$ for $1^\circ$ circular patches, averaged over a scan circle. 
On average, the ZLE contribution is 
largerly above the instrumental noise from 857~GHz to 353~GHz.
 
\subsection{Time dependence characterization}\label{sec:time:dependence}

The time dependence in the ZLE signal is characterised by the short term modulation
shown in Fig.~\ref{Fig1} and by a long term modulation derived from the effective 
motion of the spacecraft within the IDP cloud determined by the \LT\ orbital motion
around the Sun and the spacecraft Lissajous orbit around \LT.

The effects of the spacecraft motion are better represented in the reference frame 
defined by the cylindrical symmetry of the cloud and corotating with \LT. 
To understand the long term modulation we have to consider the followings:

\begin{enumerate}

 \item
The tilt of the IDPs fundamental plane over the ecliptic, which
introduces a seasonal modulation in the signal seen by a
spacecraft bound stay near \LT;

 \item
The ellipticity of the \LT\ orbit about the Solar System
barycentre moving the spacecraft with respect to the centre of the
cloud;

 \item
Since the centre of the Smooth component does not coincide with the
Sun, even a circular orbit around the Sun will induce changes in
the spacecraft position with respect to the cloud centre of symmetry;

 \item
Following its Lissajous orbit around \LT\ the spacecraft changes
its height over the cloud symmetry plane.

\end{enumerate}

The spacecraft position is then $\SpaceCraft = \SpaceCraftLT +
\DSpaceCraftLJ$, with $\SpaceCraftLT$\ the position of the \LT\ 
point and $\DSpaceCraftLJ$ the Lissajous orbit.
For typical \Planck\ Lissajous orbits around \LT, $|\DSpaceCraftLJ|$ is less 
than a few $\times 10^5$~Km.

Due to the ellipticity of the Earth orbit, the distance of \LT\ from the Sun varies
during the year by a 3\%, i.e. of $0.03$~AU~$\sim 4.5\times10^6$~Km.

Being on the ecliptic, the \LT\ point changes its distance from
the bulk component symmetry plane due to its tilt. This induces 
at maximum a vertical oscillation of $\pm 5.2\times
10^6$~Km.
In addition the Sun is off-centred with respect to the centre of IDP cloud 
of the Smooth component $\approx2\times10^6$~km.

The largest seasonal dependence is due to the tilt of the IDP fundamental plane. 
It affects mainly the value of the minima of the surface brightness 
observed by \Planck. 
When the spacecraft is below the fundamental plane of the IDP cloud the optical depth 
towards the North ecliptic Pole is  larger  than that  towards the South ecliptic Pole, 
resulting in a North/South  asymmetry in the minima.
As \Planck\ orbits about the Sun, the spacecraft goes toward
the node between the ecliptic and the IDP cloud symmetry plane, 
crosses it and enters a region where the symmetry plane is 
below the ecliptic. So, with time the North/South
asymmetry goes to zero and then reverts its sign.
The tilt of the symmetry plane over the ecliptic does not significantly affect the 
level of maximal ZLE surface brightness observed by \Planck, while it affects the 
location of the maxima and the shape about the peak.
 Fig.~\ref{fig:odd:even:asim}\ represents the modulation of 
the minima and the North/South asymmetry for a 857~GHz horn
supposed to be aligned with the telescope optical axis.
The full curve represents the surface brightnesses looking to the North 
ecliptic
Pole.
The dashed curve represents the surface brightnesses looking towards
the South ecliptic Pole.
The relative seasonal modulation is about $20\%$.

In Fig.~\ref{fig:yearly:aver:circ} we compare the yearly averaged ZLE 
surface brightness with the daily surface brightness at 857~GHz. 
The variation of the \Planck\ position with respect to the symmetry
plane of the Smooth component introduces variations of up to $\simeq10\%$ in the
surface brightness with respect to the yearly average surface brightness, 
almost independently of the considered frequency and \Ef.  
For typical Lissajous orbits, the variation of the \Planck\ height with respect
to the cloud symmetry plane is $\sim 10\%$ of the variation induced by
the tilt of the symmetry plane on the ecliptic but with a periodicity of 
6 months and phase displacement with respect to the yearly periodicity related
to the exact launch date. 
Therefore, about 10\% of the above 10\% variations of surface brightness 
induced by the effective  
\Planck\ motion is introduced by the Lissajous orbit. Clearly, this is a second
order effect for studies of the yearly averaged properties of the Smooth
component, but it is still larger than the sensitivity of \Planck\ TODs 
averaged over $1^\circ$ or $2^\circ$ resolution, as it will be discussed in 
Sect.~\ref{sec:separability}.
Secondary components contribute to about 10\% of the global ZLE. 
Therefore, neglecting the 
\Planck\ orbit may significantly reduce the accuracy with which these components
can be studied.
Finally, the differential approach to ZLE separation that has several advantages with 
respect to other approaches (see Sect.~\ref{sub:sec:differential:fitting})
exploits the variation of the ZLE during the mission. Neglecting the
Lissajous orbit effect will result in an error of $10\%$ in this kind of analysis. 
Of course, the precise inclusion of the spacecraft position is not a concern from a 
computational nor practical point of view.

\FIGYEARLYAVERCIRC

 \subsection{Frequency scaling for the spatial distribution}\label{sub:sec:freq:scaling}

The frequency scaling for $Z_{\mathrm{f}}(\Point, \Sun, \SpaceCraft)$
is a theoretical outcome of the model and can be used to check 
the extent by which it is possible to assume 
$Z_{\mathrm{f}}(\Point, \Sun, \SpaceCraft) \propto f^{\gamma_{f,z}}$. 
To study the frequency scaling for the spatial distribution of ZLE, sets of 
$Z_{\mathrm{f}}(\Point, \Sun, \SpaceCraft)$ have been generated  
for $f$ covering
all the \Planck\ frequency channels up to $f=1200$~GHz,
a fixed combination of 
\Sun\ and \SpaceCraft\ positions, $E_f \equiv 1$, 
and scanning the sky in circles of increasing angular radius from $75^\circ$ to 
$95^\circ$ centred in antisolar directions.
The corresponding data are plotted in the upper frame of 
Fig.~\ref{Fig:Beta:Plot},
while the lower frame represents the spectral index obtained
by fitting a power law dependence for the surface brightnesses obtained 
for a given
pointing direction.

Note that the average $\gamma_{f,z}$ is $\approx 1.971$, close to the expected 
value $\gamma_{f,z} = 2$. 
In addition, the spectral index is modulated with the pointing and is anticorrelated 
with the surface brightness, the higher spectral indices occurring for  
lower surface brightnesses. 
The amplitude of the modulation, however, is modest, 
$\Delta \gamma_f \approx 9.4\times10^{-3}$ 
in terms of peak-to-peak signal.
The colour correction would be then insignificantly affected.
The effect of a change of $\pm 0.07$~AU in the position of \Planck\
over the ecliptic results
in a $\delta \gamma_{f,z} < 0.002 \gamma_{f,z} \lsim 0.004$ which has
negligible effects on \Kf\ too. Smaller variations occur shifting the 
\Planck\ position in other directions. 
The same holds for a shift of the Sun position with respect to the centre of 
the cloud.

These results assure that a power law is an adequate
approximation within each frequency band, but this is only
approximately true considering the full range of \Planck\ frequencies. 
For example, in the range $70 \; \mathrm{GHz} \le f \le 144 \; \mathrm{GHz}$ 
the mean spectral index is 1.990 while in the range 
$545 \; \mathrm{GHz} \le f \le 1000 \; \mathrm{GHz}$  the mean spectral index is 1.928. 

 Assuming at $f = 100$~GHz the same spectral index 
 as at $f=857$~GHz,
 the relative error induced in the surface brightness prediction at 
100~GHz
 would be $\approx 7\%$. 
Therefore, a set of spatial templates, 
one for each frequency channel, has to be 
produced in order to simulate the spatial dependence of
the ZLE with an accuracy of $1\%$. 
On the contrary, by relaxing the required accuracy to $10\%$, 
a spatial template at an appropriate reference 
frequency (for example at $f = 1000$~GHz) 
followed by a spatially independent frequency scaling can be used.

These effects, although very small, may be not negligible in CMB studies, that are
mainly carried out at $\nu \sim 100$~GHz requiring foreground removal with an accuracy 
better than $\simeq 1\%$.

\TABEFRANDOMPOINTING

 \section{Separability of ZLE with \Planck\ data}\label{sec:separability} 

The ZLE represents a foreground contamination mainly relevant for the 
higher frequency channels of
\Planck. 
Its contribution has to be removed to accurately study the Galactic
large scale structure and its frequency scaling.
Given the weakness of the ZLE signal, the final separation quality 
will may rely on the prior information added to the system  
derived from other missions at IR bands, such as IRAS and COBE where the
ZLE dominates the sky emissivity. 
We discuss here four different approaches to ZLE separation.

 \subsection{A test of a ``blind'' map-based approach}

When ZLE is folded over a map, its histogram is strongly
non-Gaussian, as for the Galaxy.
In principle this suggests the possibility of obtaining a proper
separation of the ZLE through blind component
separation methods already used to analyze microwave maps. 

We performed some numerical experiments with the {\tt FastICA} code
\citep{FastICA} applied to full-sky maps obtained by adding the ZLE and 
the Galactic emission. {\tt FastICA} is a blind-separation
method which usually uses as input $N_{\mathrm{maps}}$ maps at 
different frequencies that are linear combinations of 
$N_{\mathrm{signals}} (\le N_{\mathrm{maps}})$
signals, all non-Gaussian except for at most one.
The code gives as output maps of the various signals.
We exploited the two frequency channels at 857~GHz and 545~GHz 
and, as a test case, neglected the noise.
Although in principle this method could be investigated, possibly by applying it to 
\Planck\ channels combined with IR data, this test gives discouraging results.
Likely this is due to the weakness of
the ZLE emission and the fact that this code does not use any 
prior information about the ZLE spatial distribution. In addition 
part of this information is lost when passing from TODs to maps.

 \subsection{A ``non-blind'' map-based approach}\label{sub:sec:nonblind:map}

An ``ad hoc'' strategy for ZLE detection and separation can use a prior information, 
derived from IR observations, on the spatial dependence of the ZLE.
We developed a model for ZLE detection and separation based on the extrapolation of the 
geometrical information from COBE/DIRBE to \Planck\ frequencies (or from any other 
reliable model), leaving as free parameters the emissivity corrections, $\Ef$, at 
\Planck\ frequencies.
The ability of \Planck\ to measure the ZLE is then 
translated into its accuracy in the determination of \Ef\ at different
frequencies.

A non-blind separation based on maps could be differently investigated
taking as prior information the ZLE spatial dependence and 
the existing templates of the Galactic emission degraded to angular
resolutions comparable with the scales of significant ZLE 
variations ($\approx 5^\circ - 10^\circ$).

For example, denoting with 
 $\mathcal{S}_f$ the map obtained from the observed signal minus its  
 average, $\mathcal{G}_f$ the Galactic emission template (or any other 
relevant background signal) minus its average, 
 $\mathcal{Z}_f$ the template for ZLE spatial distribution minus its  
 average  and
 $\mathcal{N}_f$ the template for noise spatial distribution minus its  
 average for the 
 considered scanning strategy and satellite orbit, then 
 the map for the signal (minus its average) may be approximated with

  \begin{equation}\label{eq:signal:map}
 \mathcal{S}_f \simeq E_f \mathcal{Z}_f + \alpha_f \mathcal{G}_f + \mathcal{N}_f \;,
  \end{equation}

 \noindent
where $\alpha_f$ 
accounts for possible overall systematic 
scaling errors,
both from calibration and frequency extrapolation, in the 
``Galactic'' template. 
It is possible to attempt a minimisation of

 \begin{equation}\label{eq:chisq:map}
 \chi^2 = \sum_p \frac{1}{\sigma^2_{f,p}}
 \left(
  \mathcal{S}_{f,p} - 
 \ExtEf \mathcal{Z}_{f,p} -
 \tilde{\alpha}_f \mathcal{G}_{f,p} 
 \right)^2 \; ,
 \end{equation}

 \noindent
where the sum is taken over all the pixel index of the map $p$
and $\sigma^2_{f,p}$ is the noise variance at each pixel.

Assuming stationary noise equidistributed over the map, 
the $\chi^2$ minimisation provides 
the estimators for \Ef\ and \alphaf\ 
  
 \begin{equation}\label{eq:ef:alphaf:ext:map}
 \begin{array}{lcl}
 \ExtEf & = & \;\;\; \frac{1}{\Delta} \left( \Sigma_{GG}\Sigma_{ZS} - \Sigma_{ZG}\Sigma_{GS} \right) \;, \\
 \Extalphaf & = & -\frac{1}{\Delta} \left( \Sigma_{ZG}\Sigma_{ZS} - \Sigma_{ZZ}\Sigma_{GS} \right) \;, \\
  \Delta & = & \Sigma_{ZZ} \Sigma_{GG} - (\Sigma_{ZG})^2 \;.\\
 \end{array}
 \end{equation}

 \noindent
Here $\Sigma_{ZZ} = \sum_p \mathcal{Z}_{f,p} \mathcal{Z}_{f,p} / \sigma^2_{f,p}$,
 $\Sigma_{GG} = \sum_p \mathcal{G}_{f,p} \mathcal{G}_{f,p} / \sigma^2_{f,p}$,
 $\Sigma_{ZG} = \sum_p \mathcal{Z}_{f,p} \mathcal{G}_{f,p} / \sigma^2_{f,p}$,
 $\Sigma_{ZS} = \sum_p \mathcal{Z}_{f,p} \mathcal{S}_{f,p} / \sigma^2_{f,p}$,
 $\Sigma_{GS} = \sum_p \mathcal{G}_{f,p} \mathcal{S}_{f,p} / \sigma^2_{f,p}$.

In addition, it is possible that the true scaling factor is not constant over the sky. 
To simulate this effect we replace a constant scaling \alphaf\ with a normally 
distributed variable, with expectation \alphaf\ and RMS $\delta \alpha_f$.

We consider here an illustrative case with sensitivity per pixel 
$\sigma_{f,p} \simeq 4\times10^{-3}/\sqrt{N_{\mathrm{eff},f,p}}$~MJy/sr, 
where $N_{\mathrm{eff},f,p}$ the effective number of observations
made during the mission which contributes to the pixel $p$,
$\Ef = 0.65$, $\alpha_f = 1.2$, $\sigma_{\alpha,f} = 0.01$ 
at $f=857$~GHz, and a map sampled at $\simeq 2^\circ$ resolution,
and taking all the pixels in the map (i.e. including also 
regions where the Galaxy largely dominates)
the RMS for $\Extalphaf$ is $\approx 2 \times 10^{-4}$ 
with a bias of the same order,
while $\Ef$ is recovered with an RMS accuracy of about $0.04$ but an
excess bias of about $0.06$.
On the other hand, removing all the pixels where the signal from the Galaxy does not 
greatly exceed that of ZLE reduces the bias. 
Removing pixels for which the Galaxy exceeds the surface brightness of
$4$~MJy/sr, \ExtEf\ and \Extalphaf\ are recovered with a RMS accuracy
of about 0.01 for \Extalphaf\ and 0.05 for \ExtEf. Their expectations 
are very close to their input values, with biases of a few $\times 10^{-3}$.
(In the remain we consider limits of surface brightness at values exceeding 
$1$~MJy/sr). As can be seen, the accuracy of the method is very good.

On the other hand, any (positive or negative) 
residual contribution from the ZLE in the Galactic template from
the data analysis of the IR data will be scaled to 
\Planck\ frequencies and will introduce a systematic effect which 
will be correlated with the spatial template adopted in 
Eq.~(\ref{eq:chisq:map}). This will result in biases in the recovered
\ExtEf\ and \Extalphaf\ values. An end-to-end evaluation of this
effect, beyond the scope of this paper,
would require to analyse in detail the mission and the 
data reduction procedure used to obtain each IR data set used in
preparing the Galactic template. 
However, the results of this approach can be compared with the 
differential method described in Sect.~\ref{sub:sec:differential:fitting} 
that automatically by-passes this problem.

\TABEFCALIBERR

 \subsection{A ``total-power'' TODs-based approach}
\label{sub:sec:direct:fitting}
The prior information derived from IR observations, discussed previously also can be 
applied to the time domain, taking also into account the time dependence of the 
spacecraft position within the IDP cloud.

In this approach the separation is based on the knowledge of the time dependence of the 
ZLE signal in TODs derived from the spatial pattern $Z_f(\Point{},\Sun,\SpaceCraft{})$. 
Then, as before, we define an estimator $\ExtEmEff{f}$\ of $\EmEff{f}$\ starting from 
the observed data and the known spatial pattern.

Again, denoting with $\Galaxy{t}$, $\Zodiacal{t}$,  $\Noise{t}$, and
$\Signal{t}$ the Galaxy, the ZLE, the noise and the signal
(minus their average values over the mission), and neglecting systematic
instrumental effects 

 \begin{equation}\label{eq:signal:model}
 \Signal{t} = \Ef \Zodiacal{t} + \alpha_f \Galaxy{t} + \Noise{t} \;.
 \end{equation}

 \noindent
This equation is analogous to Eq.~(\ref{eq:signal:map}) where $p$ is 
replaced by $t$ and maps are replaced by TODs. So \ExtEf\ and 
\Extalphaf\ can be obtained from Eq.~(\ref{eq:ef:alphaf:ext:map}) 
but replacing $p$ with $t$ and maps with TODs. 
Note that in the case of 
stationary noise~\footnote{We consider in this work instrumental white 
random noise. {\sc Planck} receivers are affected by $1/f$-like noise 
that introduces long term correlations appearing as offsets in TODs 
(and as stripes in maps). On the other hand, destriping algorithms 
\citep[see, e.g., ][]{Burigana:etal:1997} 
accurately remove this effect in 
the TODs (and in the maps).}, 
uncorrelated with the signal,
 $\sigma_t$ is constant all over the TOD and
substitutions like $\Sigma_{ZG} = \sum_t Z_t G_t / \sigma_t^2 = \cov(Z,G)/\sigma_f^2$ are allowed 
 \footnote{This is not true for a map since in this case $\sigma_p^2$ 
 is a function of $p$ even for stationary white noise.}.
The $\chi^2$ surface in this case is similar to the case in 
Sec.~\ref{sub:sec:nonblind:map},
but here the sensitivity per
pixel is constant $\sigma_{f,t} \simeq 4\times10^{-3}$~MJy/sr.
The final sensitivity of this approach is found to be similar to that
found for the method in Sec.~\ref{sub:sec:nonblind:map}.

Consider the case in which the Galactic contribution is neglected in the fitting,
as in \citet{Kelsall:etal:1998}. Assuming stationary white noise, after 
minimisation of

 \begin{equation}\label{eq:chisq:direct:II}
 \chi^2 = \sum_{t} \frac{1}{\sigma_{f,t}^2}\left( \ExtEmEff{f} \Zodiacal{t} - \Signal{t}
 \right)^2 \; ,
 \end{equation}

 \noindent
the estimator formula is 
 
 \begin{equation}\label{eq:extimator:direct}
    \ExtEmEff{f} = \frac{\SumSZ}{\SumZZ} \; .
 \end{equation}
  
Assuming that the noise is uncorrelated with the sky signal, the 
expectation for $\ExtEmEff{f}$\ is

 \begin{equation}\label{eq:extimator:direct:expectation}
    E[\ExtEmEff{f}] = \EmEff{f} +
    \frac{\cov(\Galaxy{},\Zodiacal{})}{\var({\Zodiacal{}})} \; .
 \end{equation}

 \noindent
This simple estimator is then affected by a bias due to the
correlation between \ZodMod\ and \GalPat.

The bias is likely negligible for \citet{Kelsall:etal:1998}
since in their case the ZLE signal is much larger than the Galaxy
emissivity so that $\left|\cov(\Galaxy{},\Zodiacal{})\right| \ll
\var({\Zodiacal{}})$.
But at 857~GHz we obtain
$\cov(G,Z) \approx 0.16$~MJy$^2$/sr$^2$, to be compared with 
 $\var({\Zodiacal{}}) \approx 0.047$~MJy$^2$/sr$^2$
 and
 $\var({\Galaxy{}}) \approx 1117$~MJy$^2$/sr$^2$,
 leading to a bias $\approx 3.4$ in $\EmEff{f}$.
Such high bias such a small covariance comes from the fact that
$\cov(\Galaxy{},\Zodiacal{})$\ is of the same order of magnitude
as $\var({\Zodiacal{}})$.
Selection of samples in order to reduce
$\left|\cov(\Galaxy{},\Zodiacal{})\right|/ \var({\Zodiacal{}})$
does not mitigate the problem. 
For example, removing all the samples where the Galaxy is larger
than $1$~MJy/sr leads to 
$\cov(\Galaxy{},\Zodiacal{}) \approx 7.6\times 10^{-3}$~MJy$^2$/sr$^2$, 
$\var({\Zodiacal{}}) \approx 0.033$~MJy$^2$/sr$^2$,
$\var({\Galaxy{}}) \approx 0.034$~MJy$^2$/sr$^2$, 
with a bias $\approx 0.23$.
Removing all samples where the Galaxy is larger than
$0.4$~MJy/sr leads to 
$\cov(\Galaxy{},\Zodiacal{}) \approx 7.6\times 10^{-4}$~MJy$^2$/sr$^2$, 
$\var({\Zodiacal{}}) \approx 5.5 \times 10^{-3}$~MJy$^2$/sr$^2$,
$\var({\Galaxy{}}) \approx 2.4 \times 10^{-3}$~MJy$^2$/sr$^2$, 
with a bias $\approx 0.14$. 
As evident, the bias decreases when applying stronger cuts but it still remains 
significant (see Table~\ref{tab:gal:zle:corr}, columns 1 to 6). 

 \subsection{A ``differential'' TODs-based approach}
\label{sub:sec:differential:fitting}

\Planck\ will scan the sky at least twice during the mission.
Therefore, most of the sky directions will be observed at least twice
with \Planck\ in different positions within the IDP cloud. 
In the ideal case, the difference between these two measures will be 
due to the difference in the ZLE contribution that can be predicted 
from our model plus noise.

We denote with $\tFirst(\Point)$ and $\tSecond(\Point)$ the epochs of 
the first and the second observation of a region seen in the direction 
\Point\ and with 
 $S_{\mathrm{I}}$, $S_{\mathrm{II}}$ 
 ($Z_{\mathrm{I}}$, $Z_{\mathrm{II}}$ or 
  $G_{\mathrm{I}}$, $G_{\mathrm{II}}$ or 
 $N_{\mathrm{I}}$, $N_{\mathrm{II}}$)
the corresponding observed surface brightness total signal variations (ZLE spatial 
distribution or Galactic emission or noise) with respect to the mean. With these 
definitions the differential surface brightnesses will be

 \begin{equation}\label{eq:signal:model:delta}
 \DeltaS{t} = 
   S_{\tSecond} - S_{\tFirst} = 
 \DeltaG{t} + \EmEff{f} \DeltaZ{t} + \DeltaN{t}
 \end{equation}

 \noindent
with the convention that $t = \tFirst$ and that 
$\DeltaG{t} = G_{\mathrm{II}} - G_{\mathrm{I}}$ and so on.
The $\chi^2$ is defined now as

 \begin{equation}\label{eq:chisq:direct}
 \chi^2 = \sum_{t} 
          \frac{
          \left( \ExtEmEff{f} \DeltaZ{t} - \DeltaS{t}
          \right)^2}
          {\sigma_{\Delta N,t}^2};
 \end{equation}

 \noindent
where $\sigma_{\Delta N,t}$ is the RMS of the noise for the considered
samples 
(for stationary noise $\sigma_{\Delta N,t} = \sigma_{\Delta N,\tFirst}^2+\sigma_{\Delta N,\tSecond}^2 = 2\sigma_{\Delta N}^2$)
giving

 \begin{equation}\label{eq:extimator:difference}
    \ExtEmEff{f} = \frac{\SumDSDZ}{\SumDZDZ}.
 \end{equation}

 \noindent
where 
 $\SumDSDZ = \sum_t \Delta S_t \Delta Z_t /\sigma_{\Delta N,t}^2 \propto \cov(\Delta S, \Delta Z)$
and
 $\SumDZDZ = \sum_t \Delta Z_t ^2/\sigma_{\Delta N,t}^2\propto \var(\Delta Z)$.
The expectation for this estimator is 

 \begin{equation}\label{eq:extimator:double:expectation}
    E[\ExtEmEff{f}] = \EmEff{f} +
    \frac{\cov(\DeltaG{},\DeltaZ{})}{\var({\DeltaZ{}})} \, ;
 \end{equation}

 \noindent
since, by definition, $\Point(\tFirst) = \Point(\tSecond)$, we have
$\DeltaG{} \equiv 0$ for any pair of \tFirst, \tSecond\ giving

 \begin{equation}\label{eq:extimator:nobias}
    E[\ExtEmEff{f}] = \EmEff{f},
 \end{equation}

 \noindent
without any bias and without the need to use Galactic templates. 

Applying the standard error propagation formula to 
Eq.~(\ref{eq:extimator:nobias})
and considering Eq.~(\ref{eq:extimator:difference}), 
if the noise $\DeltaN{t}$\ can be approximated as stationary,
independent and Gaussian with 
variance $\sigmaDN_{,t}^2 \equiv \sigmaDN^2$, 
summing over all of the pairs we have 
 $$
      \var\left[\frac{\SumDSDZ}{\SumDZDZ} \right] =
      \frac{\sum_t \expect[2\DeltaN{t}^2] \DeltaZ{t}^2}{\SumDZDZ} =
      \frac{\sum_t \sigma_{\Delta N,t}^2 \DeltaZ{t}^2}{\SumDZDZ} \, .
 $$

 \noindent
After some algebra we have 

 \begin{equation}\label{eq:estimator:double:variance:wn}
 \var[\ExtEmEff{}] \approx
     \frac{2\sigmaDN^2}{\Ncpl \var(\DeltaZ{})},
 \end{equation}

 \noindent
where \Ncpl\ is the number of independent $(\tFirst, \tSecond)$ pairs. 
Note that $\var[\ExtEmEff{}]$ does not significantly depend on the adopted 
radius of the patch because of the invariance of the $\sigmaDN^2/\Ncpl$ ratio.

Fig.~\ref{fig:zle:diff} represents the expected 
$\Ef \DeltaZ{}$ for $\Ef = 0.65$, 
the nominal scanning strategy of \Planck\ at 857~GHz, 
and a set of selected ecliptical longitudes of \Planck\ 
 \footnote{
  In this calculation it is assumed that \Planck\ is orbiting about the 
L2 point according to the nominal orbit expected for a launch in February 
2007. Changes in this orbit will only slightly change the results 
discussed above.}.
Note that the peak differential signal is about $10^\circ - 20^\circ$ 
from the ecliptic plane.
The signal is calculated averaging over independent circular patches of 
$2^\circ$ in radius. 
The difference between the first and the second scan  never exceeds 0.06~MJy/sr, 
then being a $\approx 10\%$ effect. However, when compared to the sensitivity 
expected in this frequency channel (represented at $5\sigma$ by the gray band),
this signal is clearly detectable, particularly
when the spacecraft is located at ecliptical longitudes of 
$\sim45^\circ$ and $\sim270^\circ$, where a particularly good peak $S/N 
(\simeq 40)$ is expected. 
For a $S/N$ threshold $\approx 5$, a clear detection of 
the differential signal is expected for $\Ef \gsim 0.06$. 
In case of $\Ef \approx 0.2 - 0.5$, the $S/N$ ratio is so good as to open the 
possibility to also improve the parameters of the geometrical model, to study 
possible spatial dependences in $\Ef$, and to identify secondary components.

The noise statistics ``per patch'' (RMS and patch--to--patch correlation) depend on the 
method used to assemble samples from TODs to form patches of sky.
To determine a simple statistic for the noise, we construct patches
 \first\ of fixed solid 
angle (namely, circular patches with radius, $r_{\mathrm{dsc}}$,
 of $1^\circ$ or $2^\circ$), 
 \second\ observed in both surveys with a significantly large number of 
samples coadded so as to avoid a significant difference
in the effective weight of each sky direction in the two surveys
(in order to assure a similar coverage of the same 
patch in the two surveys and smooth out possible particularly bright 
pixels -- see also the discussion in Sect.~\ref{subsubsec:samp}),
 \third\ taken contiguously in time ($\Delta t \leq 1 \;\mathrm{day} \times(r_{\mathrm{dsc}}/1^\circ)$) in each survey, and  
 \fourth\ avoiding the presence of overlapping patches.
These constraints only slightly reduce the number of samples 
used in the analysis.
After two surveys the average instrumental noise RMS on a single squared pixel with 
side equal to $\BeamSize = 5$~arcmin for the reference frequency channel at 857~GHz is 43~mJy. 
Composing these pixels to form circular patches of 
radius $\DiscRadius$, the noise per patch is 
\begin{displaymath}
 \begin{array}{ll}
\sigmaDN & \simeq  43 \times 10^{-9} \mathrm{MJy} \times
    \frac{2 } {\BeamSize^2 \sqrt{\pi\DiscRadius^2 
/\BeamSize^2}} \\
 & \simeq 
 \frac{
  1.9 \times 10^{-3} }{\DiscRadius/1^\circ}
\;\mathrm{MJy/sr} \;,\end{array} \end{displaymath}
where the factor $2$ comes from the fact that we are considering the difference between 
the values observed in the same patch in the two surveys taken separately (the variance 
of samples entering in the patch can be neglected according to condition \second).

Columns 7 to 9 of Table~\ref{tab:gal:zle:corr} shows the statistics for the simulated scan 
at 857~GHz. The last column gives the expected 1~$\sigma$ error
on \Ef\ determination according to Eq.~(\ref{eq:estimator:double:variance:wn}) with the 
expected level of noise. 

Accepting all the sky samples at 857~GHz a RMS accuracy $\sigmaEf \sim 7\times10^{-4}$ 
should be expected.
The effect of cuts based on the Galactic surface brightness is shown in 
Table~\ref{tab:gal:zle:corr}.
By considering regions where the Galactic signal is smaller than 
$\Gcut = 1$~MJy/sr 
($\Gcut = 0.4$~MJy/sr) the accuracy reduces to
$\sigmaEf \sim 2\times10^{-3}$
($\sigmaEf \sim 8\times10^{-3}$). 
On the other hand, for $\Gcut = 4$~MJy/sr 
(relevant to reduce the impact of the relative calibration
uncertainty, see Sect.\ref{sec:calibration:error}) 
the number of independent pairs decreases from
$\simeq 8\times 10^3$ to $\simeq 5.1\times 10^3$ with 
$\sigmaEf \sim 10^{-3}$, only slightly degraded with respect to
the full sky analysis. 
The cut changes the sign of column~7 due to the 
tilt of the ecliptic plane, so that the ZLE and its variation is 
stronger where the Galaxy is weaker. 

A similar analysis carried out at 545~GHz and 353~GHz gives analogous
results on the scaling introduced by the surface brightness cut. 
By considering the sky regions identified by the 
$\Gcut = 4$~MJy/sr at 857~GHz, with the same kind of analysis we find 
$\sigmaEf \sim 2.1 \times 10^{-3}$ and 
$\sigmaEf \sim 2.6 \times 10^{-3}$ at 
545~GHz and 353~GHz, respectively.

 \subsection{Systematic effects in the differential approach}

The most important source of error in the determination of \Ef\ 
is the bias induced by $\cov(\DeltaG{},\DeltaZ{})$\ 
when it is comparable to $\var(\DeltaZ{})$. 
In this subsection the main sources of this bias are discussed.

 \subsubsection{Error Induced by Sky Sampling}\label{subsubsec:samp}

In the differential approach presented in the previous section we 
cancel out the Galactic signal. 
This is a good approximation provided that the patch
is equally sampled in each of the two surveys, implying $\DeltaG{}
(\Point) \equiv 0$.
In reality, uncertainties in the spacecraft attitude reconstruction
and in pointing maneuvers and, in particular, the displacements
between the positions of the various samples taken in the two surveys
will imply that the same patch is sampled in a different manner in each
survey. An estimate of the maximum displacement between the position of
a given sample in the first survey and the nearest sample taken in
the second survey is given by the maximum between half of the
spin axis displacement ($1.25$~arcmin) and half of the angular 
sampling along the scan circle ($\sim 0.5 \; \mathrm{FWHM}/3$, equivalent 
to $\sim 0.8$~arcmin at 857~GHz).
The pointing accuracy for \Planck\ is expected to be better
than $0.5$~arcmin ($1\sigma$) for each sample
\citep{pointing:req:rep:lfi,pointing:req:rep:hfi}, a value smaller than
the above 
 estimate~\footnote{
 In addition, while finalizing this paper, a significant improvement in 
 the \Planck\ star trackers was achieved leading to an expected  
 pointing accuracy of few arcsec
 (see, e.g., \citet{Harrison:2005}).
 }.
Denoting with \Point\ the pointing direction neglecting the displacements discussed 
above and with $\Point_{i}(\Point)$ the effective pointing direction in the $i$-th 
survey, the displacement is $\delta \Point_i(\Point) = \Point_{i}(\Point) - \Point$. 
We are interested in $\DeltaG{}(\Point)$ which depends on the combined displacements
$\DeltaP{}(\Point) = \delta \Point_{\mathrm{II}}(\Point) - \delta \Point_{\mathrm{I}}(\Point)$.
Then $\DeltaG{}(\Point) \simeq \Grad G(\Point) \cdot \DeltaP{}(\Point) \neq 0$, 
leading to a bias in the determination of \Ef\ of the order of 
$\cov(\Grad G \cdot \DeltaP,\DeltaZ{})/\var(\DeltaZ{})$.

If we assume that $\delta \Point_i$ does not correlate with the Galactic
emission or the ZLE, as is reasonable, no bias will be introduced.
On the other hand, this effect will degrade the sensitivity by
 \begin{equation}
 \sigma_{\Ef,\mathrm{point}}^2 = 
 \frac{
  \sum_i (\Delta Z_i)^2 \sigma_{\mathrm{D},G,\mathrm{patch},i}^2
  }{
  (\sum_i (\Delta Z_i)^2)^2
  } \; ,
 \end{equation}

 \FIGODDEVENASIM

 \noindent
where the variance $\sigma_{\mathrm{D},G,\mathrm{patch}}^2$ is taken
over the considered patch.
We assume that $\delta \Point_i$ is a bivariate random variable, 
normally distributed, with null expectation 
and isotropic covariance matrix 
 $\mathbf{C}_{\mathrm{p}} = \PErrSgm^{-2}\mathrm{diag}(1,1)$, with the 
displacement variance 
$\PErrSgm^2 \sim 1 \; \mathrm{arcmin}^2$ uniform all over the sky.
Therefore, in the above linear approximation, for each patch of \Nsmp\ 
samples the averaged surface brightness difference will be normally 
distributed with 
null expectation and variance

 \begin{equation}\label{eq:var:displacement}
\sigma_{\mathrm{D},G,\mathrm{patch}}^2 \sim \frac{2}{\Nsmp} \var_{\mathrm{patch}}(\Grad G) 
\PErrSgm^{2}\; ,
 \end{equation} 

 \noindent
where the variance is taken over the considered patch and the factor 2
is due to the differential approach.
For patches of $1^\circ$ (or $2^\circ$) radius, and samples of 
$5/3$~arcmin along the scan circle and $5/3$~arcmin transversally to it,
$\Nsmp \approx 2.5\times10^3$ (or $\approx 1\times10^4$).
For a simple determination we can assume that 
  $\var_{\mathrm{patch}}(\Grad G) 
    \approx 
   \var_{\mathrm{patch}}(G) 
   / \ell_{\mathrm{smp}}^2
  $
where $\ell_{\mathrm{smp}}^2$ is the typical solid angle of the sample,
so that 

 \begin{equation}\label{eq:simple:estimate:sigma:bias}
 \sigma_{\Ef,\mathrm{point}}^2 \sim  
 \frac{
  \sum_i  \var_{\mathrm{patch},i}(G) (\Delta Z_i)^2
  }{
  (\sum_i (\Delta Z_i)^2)^2
  } 
 \frac{\PErrSgm{}^2}{\ell_{\mathrm{smp}}^2}
 \frac{2}{\Nsmp} 
\; .
 \end{equation}

 \noindent
From the simulations of Sect.~\ref{sub:sec:differential:fitting} 
for patches of $1^\circ$ (or $2^\circ$),  $\PErrSgm=1$~arcmin
and applying a cut to remove patches where the Galaxy is stronger than
1~MJy/sr, we estimated a $\sigma_{\Ef,\mathrm{point}}$
 of $6\times10^{-4}$ (or $2\times10^{-4}$).
We simulate in detail the effect of a random pointing modelled as above.
We obtain that, apart from a few spikes that are easily filtered, the perturbation 
induced by the Galaxy is in general small.
The total error is dominated by the bias until strong cuts are applied.
The total error does not change with \Ef.
The total pointing error is small.
The results are shown in Table~\ref{Tab:Ef:Random:Pointing}
reporting both the expectation and the variance of $\Delta \Ef$. 
The expectation is small while the RMS of the \Ef\ error is consistent with 
Eq.~(\ref{eq:simple:estimate:sigma:bias}). The RMS from our MonteCarlo simulation 
scales quite well with 
 $
 \sqrt{{2}/{\Nsmp}}
{\PErrSgm{}}/{\ell_{\mathrm{smp}}}
$. 
In summary, the random pointing error does not seriously limit the recovery of \Ef.

We consider here the requirement \second\
of Sect.~\ref{sub:sec:differential:fitting}.
We could have a different 
number of samples $\Delta \Nsmp$ ($\ll \Nsmp$) in the same patch 
in the two surveys both because of the different samples at the boundary 
of the patch and because of the result of the effective scanning
strategy. 
This implies a difference in the average Galactic 
signals obtained in the two surveys, related to the fluctuations of the 
Galaxy within the patch. 
With a simple algebra it is straightforward to derive that in this case 
the variance of the induced $\Delta \Galaxy{}$ in our differential 
approach is 

 \begin{equation}\label{eq:var:boundary}
  \sigma_{\mathrm{B},G,patch}^2 \sim 
  \var_{\mathrm{patch}}(G) 
       \frac{\Delta N_{\mathrm{smp}}}
            {N_{\mathrm{smp}}} 
 \; .
 \end{equation}

 \noindent
Eqs.~(\ref{eq:var:displacement})~and~(\ref{eq:var:boundary}) clearly show the relevance 
of the constraint \second\ of Sect.~\ref{sub:sec:differential:fitting} in the 
construction of the patches.
An approximate comparison between Eq.~(\ref{eq:var:displacement}) and 
Eq.~(\ref{eq:var:boundary}) gives  $\sigma_{\mathrm{B},G,patch}^2 / 
\sigma_{\mathrm{D},G,patch}^2 \approx 
\ell_{\mathrm{smp}}^2 \Delta \Nsmp / 2\PErrSgm^{2}$. 
As $\ell_{\mathrm{smp}} > \PErrSgm$, we have
$\sigma_{\mathrm{B},G,patch}^2 \gg  \sigma_{\mathrm{D},G,patch}^2$
even for $\Delta \Nsmp \approx 1$. 
This implies that the effect associated with $\Delta N_{\mathrm{smp}}$
could significantly impact the final result,
if not taken properly into account
in the data analysis.
Of course, it has no physical meaning to have a sensitivity degradation 
in the presence of an increased number of observations in one 
of the two surveys. In reality, the above computation 
underlines the relevance of properly assembling the ``elementary'' 
samples in the two surveys in order to count the signal from
the same sky direction the same number of times in each of the 
two averages over the considered patch.

The worst effect of systematic pointing errors would occur in sky regions 
with bright point-sources. 
It would then be preferable to remove pixels affected by bright sources 
before of the computation of the averages of the signals in each patch
in order to manage only with signals dominated by the diffuse 
components.

An example of an intrinsic source of systematic pointing errors is the aberration of 
light. When not accounted for, the aberration due to satellite motion about the Sun
may induce at most a pointing error of $\approx 2 v/c$~rad~$\sim$~0.7~arcmin, dominated 
by the Earth motion about the Sun. The factor of 2 comes from the fact that patches 
are acquired at most at about $180^\circ$ of longitude when the orbital motions
are toward opposed directions in the sky. 
This effect may introduce an important bias; however since the spacecraft velocity is 
known within $1$~Km/sec or better, the effect may be removed by correcting the selected 
pointings.

 \FIGBETAPLOT

 \subsubsection{Doppler shift}

The relative motion of the satellite with respect to the Sun induces a Doppler shift in 
the Galactic signal observed during each of the two scans.
The effect will be $\delta f / f \approx 2 v/c \times 10^{-4}$ and assuming
$G \propto f^{\alpha}$ the surface brightness variation will be $\delta G 
/ G \approx \alpha 2 v/c$. 
Even for $\alpha$ in the range $2 - 3$, $|\delta G / G| \lsim 8 \times 10^{-4}$. 
Assuming $\Delta G \approx \delta G$ and taking the statistics from the
signal correlations from Table~\ref{tab:gal:zle:corr}, then 
$\cov(\DeltaG{}, \DeltaZ{}) \lsim \alpha 2 (v/c) \cov(G{}, \DeltaZ{}) \approx 1.6 \times 10^{-5}$~MJy$^2$/sr$^2$ equivalent to a bias $|\Delta \Ef| \lsim 2.6 \times 10^{-2}$. 
However the application of 4~MJy/sr cut will reduce this bias by an order 
of magnitude. Of course, a further relevant reduction 
(by a factor $\sim 10$ or 100) of this effect
can be reached with a simple modelling of the Galatic emission spectral 
index (for example at $\sim 10$\% or 1\% accuracy).

In addition, assuming the IDP cloud to be at rest around the Sun (neither shifting, nor 
rotating in time), the Doppler shift will affect the ZLE surface brightness too, so 
that a $|\Ef\DeltaZ{}| \lsim \Ef Z 2 v/c \approx 1.8 \times 10^{-4}$~MJy/sr equivalent 
to a bias in \Ef\ of at most $10^{-4}$.

 \subsubsection{Calibration uncertainty}\label{sec:calibration:error}

The impact of any absolute calibration error of \Planck\ data uniform all over the 
mission is of simple evaluation. 
In the differential approach, any calibration uncertainty of this kind produces a null 
effect in the Galactic signal.
The only final error will be a multiplicative uncertainty in the ZLE
and then in \Ef\ given by 
  $\Delta E_{f, \mathrm{g}} = \Ef \left|\CalibRErr\right| $,
where \CalibRErr\ is the relative uncertainty of the absolute 
calibration ($1\% - 3\%$ according to FIRAS absolute calibration 
accuracy from $\sim 300$~GHz to $\sim 900$~GHz).

The evaluation of the impact of calibration errors varying during the mission is more 
difficult (i.e. relative calibration errors).
Clearly, the difference in the calibration errors in each patch in the 
two considered surveys produces systematic effects in our differential 
approach proportional to 
$\CalibRErr \; \cov(G,\DeltaZ{})/\var(\DeltaZ{}) $.

We evaluated the implications of random relative calibration errors. 
The relative calibration accuracy of \Planck/HFI highest frequency 
channels is currently under definition.
\citet{Pajot:etal:2000} (\S~3.3.6.1) reported a preliminary 
relative pixel-to-pixel calibration accuracy of $\sim3\%$.
\citet{Piat:etal:2002} investigated HFI calibration with a kinematic dipole at 
frequencies at $\nu \lsim 400$~GHz and with Galactic templates at $\nu \gsim 400$~GHz. 
The authors reported
a relative calibration accuracy on each scan circle of $\sim10\%$ at 
545~GHz and of $\sim3\%$ at 353~GHz with suitable choices of the surface 
brightness cut.

 \FIGZZEROCOEF

Rescaling the error estimate by \citet{Pajot:etal:2000} from $\simeq5$~arcmin 
pixels to circular patches of $\simeq2^\circ$ radius, we find a relative accuracy, 
$\ErrPatch$, of $0.06\%$. 
The results of 
\citet{Piat:etal:2002} 
can be rescaled to the above circular patches considering that we have
about $10^2$ scan circles per circular patch. In this way we estimate
$\ErrPatch \sim 1\%$ at 545~GHz and $\ErrPatch \sim 0.3\%$
 at 353~GHz. 

\FIGZLEDIFF

 %
We then consider \ErrPatch\ in the range $0.1\% - 1\%$, so perturbing the simulated 
signal from each patch.
The comparison between the estimation of 
\Ef\ obtained in the absence of this
systematic error and by repeating the perturbed simulation described
above for many thousands of realizations and for various Galactic cuts
is shown in Fig.~\ref{fig:calib:err} for a particular case. 
As evident, at 857~GHz, assuming $\ErrPatch \simeq 1\%$
and Galactic cuts in the range $\Gcut \sim 2 - 26$~MJy/sr 
we find a RMS error on 
\Ef\ of $\sim3\times 10^{-2}$ (with an expectation value of the 
residual bias of $\lsim 10^{-4}$). 

For $\ErrPatch \simeq 1\%$ 
the RMS errors on \Ef\ 
are $\sim2.5\times 10^{-2}$ or $\sim1.5\times 10^{-2}$
respectively at 545~GHz or at 353~GHz 
for suitable Galactic cuts.
Table~\ref{Tab:Ef:Calibration:Error} summarizes our results.
The range of surface brightness cuts appropriate for each frequency 
allowing the above sensitivity spans about one order of magnitude.
We verify that for $\ErrPatch \lsim 5\%$ 
the RMS error on \Ef\ scales linearly with \ErrPatch\ at all frequencies.

Of course, the optimal Galactic surface brightness cut should be 
identified
a posteriori according to the recovered value of \Ef\
and with a proper trade-off between the various systematic
errors.

 \section{Conclusions}\label{sec:conclusions}

We presented an analysis to predict the level of contamination from the Zodiacal Light 
Emission (ZLE) in the survey of the forthcoming \Planck\ mission and to evaluate the 
ability to extract the ZLE signal from the \Planck\ data. This signal is used 
to gain more information about the ZLE physical properties.

Our starting point is the model of \citet{Kelsall:etal:1998} for the 
ZLE  based on the COBE data describing in detail the emissivity of the IDP cloud
for wavelengths up to about 300~$\mu$m complemented by the recent 
result of \citet{Fixsen:Dwek:2002} on the spectral behaviour of the ZLE.
According to the COBE model, four components contribute to the ZLE,
the dominating smooth components on which this paper is focussed,
the Earth orbit locked ring of dust (or circumsolar ring), the trailing blob,
and three bands of dust.

With respect to other foregrounds usually considered in CMB studies,
the ZLE (as the other Solar System objects) is peculiar, 
its surface brightness depending not only on the pointing direction but also on
the instantaneous position of the observer within the Solar System.
This underlines the relevance of a study of the ZLE not only on maps but also on time 
ordered data streams. 
Taking the average yearly position of \Planck\ in the IDP dust cloud
will result in a $\sim10\%$ error in the ZLE surface brightness estimate,
while neglecting the \Planck\ orbit about \LT\ will imply a 
$\sim1\%$ error in the ZLE surface brightness estimate. 
Since the ZLE differential approach separation exploits the $\sim10\%$ variation of the 
ZLE surface brightness between the two surveys, neglecting the \Planck\ orbit will 
introduce a non-negligible systematic $\sim10\%$ error in this method. 
This will have a large impact on the study of secondary ZLE components, a natural 
extension of this work.

We have implemented the COBE model in a dedicated program that computes 
$Z_f(\Point, \Sun, \SpaceCraft)$ for a given list of values of \Point, 
\Sun, \SpaceCraft\ and a set of parameters describing the properties of 
the ZLE component for which the calculation has to be performed. 
Of them, the {\em emissivity correction} \Ef\ for each component and for each frequency 
band is the hardest to extrapolate to \Planck\ frequencies and it carries most of the 
physical information on the IDP population producing the bulk emission at these 
frequency bands. We then focused on the capability of \Planck\ to recover \Ef.
Since the \Ef\ scalings are largely uncertain, the code separately generates the 
surface brightnesses appropriate for each desired component. The outputs produced for 
the various components then can be easily combined according to the user need.

Since the TODs are large and their realization is strictly related to the effective 
scanning strategy and satellite orbit we also implemented an approximate, but accurate, 
method able to {\em compress} the \Sun, \SpaceCraft\ dependencies in TODs for the 
desired pointings in matrices of appropriate series expansion coefficients.

We investigated the possibility of obtaining a proper separation of the 
ZLE through a {\em blind component separation} method (FastICA) already 
used to analyse microwave maps, without obtaining encouraging results.
 
A {\em non-blind separation} based on maps has been investigated
taking as prior information the ZLE spatial dependence and 
the existing templates of the Galactic emission degraded to angular
resolutions comparable with the scales of significant ZLE 
variations. We considered here in detail the case of the 
HFI 857~GHz channel. In this case the main foreground is the
dust Galactic emission.
For appropriate Galactic surface brightness cuts ($\sim 1$~MJy/sr) the value of \Ef\ 
recovered with this approach is in good agreement with the input one with an RMS 
absolute accuracy $\sim 0.05$.
On the other hand, any possible residual contribution from the ZLE left in the IR 
Galactic template adopted from their original data analysis procedure will be scaled to 
\Planck\ frequencies and will introduce a systematic effect which will be correlated 
with the spatial template adopted in ZLE estimates, possibly resulting in biases in the 
recovered \Ef. A total power approach on TOD has been also investigated, providing 
similar results. These two methods require the use of low resolution Galactic 
templates.

To circumvent the need for Galactic templates and to better 
take into account the effects introduced by the varying spacecraft
position, we have studied a differential approach exploiting the
fact that \Planck\ will scan the sky at least twice during the mission,
staying in different positions within the IDP cloud. 
In the ideal case, the difference between these two measures will be due to the 
difference in the ZLE contribution that can be predicted from our model plus noise.
We find a typical absolute RMS uncertainty on \Ef\
induced by the limited instrumental
sensitivity of $\sim 10^{-3}$ (1$\sigma$) at 857~GHz
for suitable choices of surface brightness cuts
(or $2.1 \times \sim 10^{-3}$ and $2.6 \times \sim 10^{-3}$ 
at 545~GHz and 353~GHz, respectively).
For typical expected values of \Ef\ 
($\approx 0.65$, 0.26, 0.11 for 857~GHz, 545~GHz, 353~GHz)
 the \Planck\ sensitivity 
will allow an \Ef\ recovery at 
0.15\%, 0.8\% and 2.4\%
(1$\sigma$) accuracy at 857~GHz, 545~GHz and 353~GHz, respectively.

We investigated the impact of the most relevant systematic effects, pointing and 
sampling uncertainty, aberration of light, Doppler shift and relative calibration 
uncertainty. While the first four effects are expected to be significantly below the 
noise, the last effect is potentially the most critical one. 
For a relative calibration error RMS of $\sim 1\%$ $(0.1\%)$ on patches of $2^\circ$ 
radius, we find an absolute RMS error on \Ef\ of $\sim 0.01 - 0.04$ 
$(\sim 0.001 - 0.004)$ with only a weak dependence on the frequency in the range 
$\sim 300 - 900$~GHz, 
corresponding to relative
errors on \Ef\ $\sim 4\%$, $10\%$, $23\%$ 
($\sim 0.4\%$, $1\%$, $2\%$) respectively at 
857~GHz, 545~GHz, 353~GHz for the most likely \Ef\ values 
expected on the basis of COBE/FIRAS data.
This may ultimately determine the final accuracy of the 
ZLE extraction from \Planck\ data. 

{\it 
A web page containing documentation, {\tt FITS} files, and 
{\tt IDL} routines to handle the series expansion for ZLE simulations
is in preparation. 
}

\FIGCALIBERR

\section*{Acknowledgements}
We warmly acknowledge the \Planck\ collaboration teams for
having provided us with instrument and mission details. We thank 
 F.~Boulanger and 
 M.~Juvela
 for constructive discussions
 and J.P.~Bernard, G.~Cremonese, M.~Fulle, and G.~De~Zotti 
 for encouraging and stimulating conversations.
We acknowledge L.~Abrami and C.~Doz of the INAF/OAT
for having supported the acquisition of bibliographic material.
Some of the results in this paper have been derived using the 
\HEALpix\ package \citep{Gorsky:etal:2005}.
We warmly thank the referee, W. T. Reach, for constructive comments 
and suggestions.
M.M. acknowledges partial support by COFIN 2005 SubMM 
(prot.2004028417\_003) SISSA / Trieste / Italy.

 
%

%


\end{document}